\documentclass[twocolumn,showpacs,aps,prd,nobibnotes,nofootinbib,floatfix]{revtex4-1}

\usepackage{amsmath}
\newcommand{\ve}[1]{\mbox{\boldmath $#1$}}
\newcommand{\beqn}{\begin{eqnarray}}
\newcommand{\eeqn}{\end{eqnarray}}
\newcommand{\beq}{\begin{equation}}
\newcommand{\eeq}{\end{equation}}
\usepackage{graphicx}
\usepackage{xcolor}
\usepackage[normalem]{ulem}
\usepackage{verbatim}

\newcommand{\IGM}{{\texttt{IllinoisGRMHD}}}
\newcommand{\Pan}{{\texttt{Pandurata}}}
\newcommand{\Whisky}{{\texttt{WhiskyMHD}}}

\newcommand{\Omfluid}{\Omega_{\rm fluid}}
\newcommand{\MSun}{M_{\odot}}
\newcommand{\Szdom}{S^z_{(1,0)}}
\newcommand{\Srdom}{S^R_{(0,0)}}
\newcommand{\valf}{v_{\rm Alf}}
\newcommand{\MTOT}{M}
\newcommand{\Mdot}{\dot{M}}
\newcommand{\LEM}{L_{\rm Poynt}}
\newcommand{\LEMinsp}{L_{\rm Poynt, insp}}
\newcommand{\LEMpeak}{L_{\rm Poynt, peak}}
\newcommand{\UNITg}{{\rm g}}
\newcommand{\UNITs}{{\rm s}}
\newcommand{\UNITG}{{\rm G}}
\newcommand{\UNITcm}{{\rm cm}}
\newcommand{\UNITerg}{{\rm erg}}
\newcommand{\umagofluid}{\zeta}

\newcommand{\rhothir}{{\rho_{\textnormal{-}13}}}
\newcommand{\press}{p}
\newcommand{\tmerge}{t_{\rm merge}}

\begin{document}
\date{\today}
\title{Prompt Electromagnetic Transients from Binary Black Hole Mergers}
\author{Bernard J.~Kelly$^{1,2,3}$, John G.~Baker$^{1,4}$, Zachariah B.~Etienne$^{5,6}$, Bruno Giacomazzo$^{7,8}$, Jeremy Schnittman$^{1,4}$}
\affiliation{$^1$ Gravitational Astrophysics Laboratory, NASA Goddard Space Flight Center, Greenbelt, MD 20771, USA}
\affiliation{$^2$ CRESST, NASA Goddard Space Flight Center, Greenbelt, MD 20771, USA}
\affiliation{$^3$ Department of Physics, University of Maryland, Baltimore County, Baltimore, MD 21250, USA}
\affiliation{$^4$ Joint Space-Science Institute, University of Maryland, College Park, MD 20742, USA}
\affiliation{$^5$ Department of Mathematics, West Virginia University, Morgantown, WV 26506, USA}
\affiliation{$^6$ Center for Gravitational Waves and Cosmology, West Virginia University, Chestnut Ridge Research Building, Morgantown, WV 26505, USA}
\affiliation{$^7$ Physics Department, University of Trento, via Sommarive 14, I-38123 Trento, Italy}
\affiliation{$^8$ INFN-TIFPA, Trento Institute for Fundamental Physics and Applications, via Sommarive 14, I-38123 Trento, Italy}

\begin{abstract}
Binary black hole (BBH) mergers provide a prime source for current and future
interferometric GW observatories. Massive BBH mergers may
often take place in plasma-rich environments, leading to the exciting
possibility of a concurrent electromagnetic (EM) signal observable by
traditional astronomical facilities. However, many critical questions about the
generation of such counterparts remain unanswered.
We explore mechanisms that may drive EM counterparts with magnetohydrodynamic
simulations treating a range of scenarios involving equal-mass black-hole binaries immersed in an
initially homogeneous fluid with uniform, orbitally aligned magnetic fields. We find that the time
development of Poynting luminosity, which may drive jet-like emissions, is relatively insensitive to
aspects of the initial configuration. In particular, over a significant range of initial values, the
central magnetic field strength is effectively regulated by the gas flow to yield a Poynting luminosity of
$10^{45}-10^{46} \rhothir{M_8}^2 \, \UNITerg \, \UNITs^{-1}$, with BBH mass scaled to
$M_8 \equiv M/(10^8 \MSun)$ and ambient density $\rhothir \equiv \rho/(10^{-13} \, \UNITg \, \UNITcm^{-3})$. We also calculate the direct
plasma synchrotron emissions processed through geodesic ray-tracing. Despite lensing effects and dynamics,
we find the observed synchrotron flux varies little leading up to merger.
\end{abstract}

\maketitle

\section{Introduction}

One of the more provocative developments associated with the recent detections of gravitational
waves (GWs) from mergers of stellar-mass black holes (BHs) by Advanced
LIGO~\cite{Abbott:2016blz,Abbott:2016nmj} was the subsequent announcement of a possible
electromagnetic (EM) counterpart signal, 0.4s after the GW150914 signal was observed.
Fermi found a sub-threshold gamma-ray source in a region of the sky that overlapped the
$\sim 600$-square-degree LIGO uncertainty region for GW150914 \cite{Connaughton:2016umz}.
Though it may be impossible to confirm that the events are indeed physically related, the
EM observation has inspired a number of papers exploring potential scenarios linking EM
counterparts to stellar-mass black hole mergers
\cite{Perna:2016jqh,Li:2016iww,Zhang:2016rli,Loeb:2016fzn,Morsony:2016upv,Murase:2016etc,Janiuk:2016qpe}---mergers
that theorists had expected to be electromagnetically dark.

These events draw attention to the high potential value of multimessenger observations of GW events.
While GW observations can provide extraordinarily detailed information about the merging black holes
themselves, they may not provide any direct information about the black holes' environment.  Even
the location of the event will be poorly determined unless an
associated EM event can be identified. Such localization could also
deepen our understanding of the astrophysical processes that form and
influence BBH systems.

Unlike the situation for stellar-mass black holes, astronomers have long recognized the potential
for EM counterparts to binary {\it supermassive} ($10^6$ -- $10^9 \MSun$) black hole (SMBH) mergers
occurring in the millihertz GW band.  These mergers are a key target of future space-based GW
observatories such as the LISA mission, which was recently approved by
the European Space Agency \cite{Audley:2017drz}. Pre-merger
GWs from these systems are a key target of nanoHertz GW searches with
pulsar timing arrays \cite{Arzoumanian:2014gja}.

The large cross-section of SMBHs interacting with the ample
supplies of gas common in galactic nuclear regions allows them to
power some of the brightest, most
long-lasting EM sources in existence, including active galactic nuclei (AGN),
quasars, or radio jet emissions. A number of mechanisms may provide
signals associated with these sources across a broad range of timescales
from $\sim 10^9$ years before merger to $\sim 10^9$ years after merger
\cite{Schnittman:2010wy}. Considerable evidence for binary SMBH systems
has already been observed, but is restricted to those either well 
before merger \cite{Komossa:2002tn,Comerford:2008gm,Smith:2009ct,Rodriguez:2006th,Valtonen:2008tx,Gabanyi:2016snk,Frey:2012pf,Yang:2016ygm},
or well after merger \cite{Merritt:2004xa,Boylan-Kolchin:2004tf,Gualandris:2007nm,Komossa:2008qd,Guedes:2009dr,Batcheldor:2010vd,Chiaberge:2016eqf}.

The greatest potential for direct association with BBH mergers would
come from strong EM emissions or modulations coincident with
the GW event. Since LISA will observe GW emission from BBH mergers for
an extended period of time, direct EM counterparts may be caused by
interaction of the BBH with a circumbinary disk, perhaps during the
final $\sim 10^3$ orbits prior to merger. Our objective, however, is
to explore the mechanisms that may potentially drive EM signals
directly associated with the strongest GW emissions within hours
of the merger event itself. Such emissions could be crucial, for example,
in LISA-based redshift-distance studies \cite{Tamanini:2016zlh}. 

Unlike the clean GW predictions that numerical relativity provides,
one challenge of understanding EM counterpart signatures is their
potential dependence on myriad details of the gas distribution, its
properties, and the structure and strength of associated magnetic
fields. For prompt merger-associated signals, the challenge is
enhanced because the merger occurs on a very short timescale.
Accretion disks, and indeed circumbinary disks, are characterized by
variation over a wide range of timescales \cite{Vaughan:2003by};
after ``decoupling'', the gravitational-radiation-induced infall
timescale becomes shorter than the disk accretion timescale, leading
to a merger in a magnetized matter environment whose detailed
structure may be impossible to predict.
Even though binary torques tend to evacuate much of the surrounding region, studies in 2D \& 3D
reveal that dense infalling streams persist, maintaining accretion rates at levels comparable to
that of a single-BH disk \cite{Farris:2014zjo,Shi:2015cjt}.

The most valuable sort of counterpart prediction would be insensitive
to these details, and have distinguishing features that clearly
identify the source as a binary SMBH. While one approach to exploring
this could be to seek universal features in a large number of full
circumbinary-disk-plus-merger simulations, our approach here is to
explore robust EM counterpart signatures from BBHs embedded in a
number of simple plasma configurations.

In this paper, we employ a new tool---the \IGM{}
code~\cite{Etienne:2015cea}---to study potential EM signals deriving from
perhaps the simplest such initial configuration: a plasma with uniform
density and magnetic fields, in which the magnetic fields are aligned
with the orbital angular momentum vector.

The rest of this paper is laid out as follows.
In Sec.~\ref{sec:hist_results}, we summarize relevant numerical results obtained with various
methods and codes.
In Sec.~\ref{sec:num_methods}, we briefly introduce our numerical code and MHD diagnostics, and
compare with results from earlier work that used the \Whisky{} code \cite{Giacomazzo:2012iv}.
Section~\ref{sec:results} presents the results of our new simulations: the global state of the
MHD fields (\ref{ssec:results_global}), the rate of mass accretion into the pre-merger and post-merger
BHs (\ref{ssec:results_mdot}), the detailed behavior of the resulting Poynting luminosity
(\ref{ssec:results_poynting}), and possible direct emission of observable photons
(\ref{ssec:results_pandurata}).
We summarize our conclusions and discuss future work in Sec.~\ref{sec:conclusions}.
The Appendices contain more detail on the calculation of the Poynting luminosity, effects of
varying numerical resolution, and conversion between code and cgs units.

\section{GRMHD simulations}
\label{sec:hist_results}

As dynamical, strong-field gravitational fields may drive EM
counterparts to GW mergers, it is essential to build our models using
the techniques of numerical relativity. Building on a revolution in methodology
\cite{Pretorius:2005gq,Campanelli:2005dd,Baker:2005vv}, numerical
relativity simulations provided the first predictions \cite{Baker:2006yw} of astrophysical
GW signals like GW150914 almost ten years
before the observation. Moving beyond GW predictions in vacuum
spacetimes and into EM counterpart predictions requires
physics-rich simulation studies that couple the general relativistic
(GR) field equations to the equations of GR magnetohydrodynamics
(GRMHD), so that magnetized plasma flows in strong, dynamical
gravitational fields may be properly modeled.

Over the last decade several research teams have gradually and
systematically added the layers of physics necessary to begin to
understand the potential for counterpart signals. Studies of test particle motion
(\emph{i.e.}~non-interacting gases) during the last phase of inspiral
and merger of MBHs showed that a fraction of particles can collide with each other at speeds
approaching the speed of light, suggesting the possibility of a burst of
radiation accompanying black hole coalescence~\cite{vanMeter:2009gu}. 
Other studies investigated possible EM emission from purely hydrodynamic fluids
near the merging BHs~\cite{ONeill:2008dg,Farris:2009mt,Bode:2009mt,Bogdanovic:2010he,Bode:2011tq,Farris:2011vx}.

These studies neglected the important role that magnetic fields are
likely to have in forming jets, driving disk dynamics, or in
photon emission mechanisms. EM fields were first included in
ground-breaking GR force-free electrodynamics (GRFFE)
simulations~\cite{Palenzuela:2009yr,Palenzuela:2009hx,Mosta:2009rr},
investigating mergers in a magnetically dominated plasma, indicating
that a separate jet formed around each BH during the
inspiral. At the time of the merger, these two collimated jets would
coalesce into a single jet directed from the spinning BH
formed by the merger~\cite{Palenzuela:2010nf,Palenzuela:2010xn,Moesta:2011bn}. 
Based on the black hole membrane paradigm, analysis of these studies suggested a simple formula
relating the binary orbital velocity to the Poynting flux available to drive EM emissions
\cite{Neilsen:2010ax}: $\LEM \sim v^2_{\rm orbital}$.

More recent studies have begun to explore the behavior of \emph{moderately} magnetized plasmas around BBH
systems in an ideal GRMHD context, finding that significant EM
signatures may be produced by these systems. Studies of circumbinary
disk dynamics \cite{Noble:2012xz,Farris:2012ux,Gold:2013zma} have
used initially circular binary BH orbits to reach a pseudo-steady state in a
circumbinary disk before allowing the binary to inspiral and merge. In
\cite{Gold:2014dta}, the final merger of an equal-mass BBH was modeled
in full GR, and the observed Poynting luminosity declined gradually
through inspiral, only to rise significantly some time after merger.

In \cite{Giacomazzo:2012iv} we first studied the physics of moderately magnetized plasmas near the
moment of merger, using the \Whisky{} code. Though that study was limited to just a few orbits
because of technical challenges, it showed a rapid amplification of the magnetic field of
approximately two orders of magnitude. This contributed to the creation, after merger, of a
magnetically dominated funnel aligned with the spin axis of the final
BH. The resulting Poynting luminosity was estimated to be 
$\sim 10^{48}\,\UNITerg \, \UNITs^{-1}$ (assuming an initial BBH
system mass of $10^8 \, \MSun$, an 
initial plasma rest-mass density of $10^{-11}\, \UNITg \, \UNITcm^{-3}$, and an initial
magnetic field strength of $\sim 10^4 \UNITG$).
In comparison, the force-free simulations of \cite{Palenzuela:2010nf,Palenzuela:2010xn,Neilsen:2010ax,Moesta:2011bn}
produced peak luminosities of $\lesssim 10^{44} \, \UNITerg \, \UNITs^{-1}$,
four orders of magnitude lower than what we obtained with ideal GRMHD, despite similar initial magnetic field strengths.

These results indicate that the dynamics of
BBH inspirals and mergers play an important role in driving the magnetic fields in their environment.
When the BBH is embedded
in an initially non-magnetically dominated plasma, accretion onto the
merging BHs compresses and twists the magnetic field
lines, which may strongly amplify the magnetic fields.
Strengthened magnetic fields may then influence gas inflow, powering a strong
EM energy (Poynting) outflow through a magnetically dominated
funnel.
As noted in \cite{Giacomazzo:2012iv}, such a mechanism cannot exist in the force-free regime.
  
Despite the ability to track GRFFE/GRMHD flows, there have been no
fully GR simulations of EM counterparts to BBH mergers that actually track photons,
or that could produce spectra. Instead, EM luminosity estimates have often
been based on Poynting flux measurements provided directly from GRMHD fluid variables.
A first step in bridging this gap was made in \cite{Schnittman:2013qxa} using the \Pan{} code
\cite{Schnittman:2013lka} to post-process the MHD fields around the merging binary, but assuming a
fixed Kerr BH background instead of the dynamical spacetime of the GRMHD evolution itself, and
also assuming a fixed electron temperature. Synchrotron, bremsstrahlung, and inverse-Compton
effects combined to produce a spectrum that peaked near 100 keV. As
described in Sec.~\ref{ssec:results_pandurata}, we apply a slightly
more sophisticated spacetime procedure with \Pan{} to obtain estimates of
synchrotron luminosity and spectra from simulations presented here.

\section{Numerical Methods}
\label{sec:num_methods}

We revisit the scenario studied in~\cite{Giacomazzo:2012iv} with fully
3D dynamical GRMHD evolutions carried out with the Einstein
Toolkit \cite{Loffler:2011ay,etk_web} on adaptive-mesh refinement (AMR) grids
supplied by the Cactus/Carpet infrastructure \cite{Schnetter:2003rb}, adopting a fully general-relativistic,
BSSN-based \cite{Nakamura:1987zz,Shibata:1995we,Baumgarte:1998te} spacetime metric evolution provided by the
Kranc-based \cite{Husa:2004ip} \texttt{McLachlan} \cite{Brown:2008sb,mclachlan_web} module, and crucially,
fluid and magnetic field evolution performed with the
recently released \IGM{} code \cite{Etienne:2015cea}.
Initial metric data was of the Bowen-York type commonly used for moving puncture
evolutions \cite{Bowen:1980yu,Brandt:1997tf}, conditioned to satisfy the Hamiltonian
and momentum constraints using the \texttt{TwoPunctures} code \cite{Ansorg:2004ds}.

The \IGM{} code is a complete rewrite of (yet agrees to roundoff-precision with) the long-standing
GRMHD code used for more than a decade by the Illinois Numerical Relativity group to model a large
variety of dynamical-spacetime GRMHD phenomena (see, e.g.,
\cite{Etienne:2006am,Paschalidis:2011ez,Paschalidis:2012ff,Paschalidis:2014qra,Gold:2014dta}
for a representative sampling).
It evolves a set of conservative MHD fields $E \equiv \{ \rho_*, \tilde{\tau}, \vec{\tilde{S}}, \vec{\tilde{B}} \}$,
derived from the primitive fields $\rho$ (baryonic density), $\press$ (fluid pressure), $v^i$ (fluid
three-velocity $\equiv u^i/u^0$, where $u^{\mu}$ is the fluid four-velocity), and $B^i$ (spatial
magnetic field measured by Eulerian observers normal to the spatial slice).

For an ideal gas with adiabatic index $\Gamma$, the pressure $\press$ obeys
\beq
\press = \rho \epsilon (\Gamma -1),
\eeq
where $\epsilon$ is the specific internal energy of the gas. The fluid specific enthalpy is
\beq
h = (1 + \epsilon) + \frac{\press}{\rho} = 1 + \Gamma \epsilon.
\eeq
More specifically, we choose the gas to initially obey a polytropic equation of state:
\beq
\press = \kappa\rho^\Gamma, \label{eq:polytrope_EOS}
\eeq
with $\Gamma = 4/3$, consistent with a radiation-dominated plasma.

We also use the magnetic four-vector $b^{\mu}$ given by (see e.g. Section II.B of \cite{Duez:2005sf}):
\beq
b^{\mu} = \frac{1}{\sqrt{4\pi} \alpha} \left( u_m B^m, \frac{B^i + (u_m B^m) u^i}{u^0} \right),
\eeq
where repeated Latin indices denote implied sums over spatial components only.
We define a specific magnetic + fluid enthalpy by
\beq
h^* = h + \frac{b^2}{\rho}. \label{eq:mag_enthalpy}
\eeq

The total stress-energy tensor of the magnetized fluid is the sum of fluid and EM parts:
\begin{subequations}\label{eq:Tab}
\begin{align}
T^{\mu \nu}             &= T_{\rm fluid}^{\mu \nu} + T_{\rm EM}^{\mu \nu}, \label{eq:Tab_sum}\\
T_{\rm fluid}^{\mu \nu} &= \rho h u^{\mu} u^{\nu} - \press g^{\mu \nu}, \label{eq:Tab_fluid}\\
T_{\rm EM}^{\mu \nu}    &= b^2 \left( u^{\mu} u^{\nu} + \frac{1}{2} g^{\mu \nu} \right) - b^{\mu} b^{\nu}. \label{eq:Tab_EM}
\end{align}
\end{subequations}

GR provides that the stress-energy tensor is equal to a multiple of the Einstein tensor, containing
information about the spacetime geometry. However, the low-density
fluids we study possess negligible self-gravity, so as
in~\cite{Giacomazzo:2012iv}, we ignore the plasma
contribution to the GR field equations. In this case we are then free
to rescale $T^{\mu\nu}\approx0$ (and thus an appropriate combination
of the plasma field variables) independently of the scaling of
geometric properties, represented by the total black hole mass
$\MTOT$.
To justify this approach more quantitatively, we note that in \cite{Fedrow:2017dpk}, the authors
found plasma densities of around $10^6 \UNITg \, \UNITcm^{-3}$ were necessary to noticeably affect
the binary's coalescence dynamics --- 17 orders of magnitude larger than the densities considered
here.

The original simulations of \cite{Giacomazzo:2012iv} were carried out
with an equal-mass binary with initial separation $d = 8.4M$, where
$M$ is the sum of Arnowitt-Deser-Misner (ADM) masses \cite{Baker:2002gm} of the pre-merger black
holes.
As reviewed in Sec.~\ref{ssec:Whisky2012comp}, in this
work we explore a variety of additional separations, better resolve
the spacetime fields near the black holes, allow for plasma
shock-heating, and adopt the new \IGM{} code for modeling the GRMHD
dynamics.

To each BBH configuration, we add an initially uniform,
radiation-dominated polytropic fluid: $\press_0 = \kappa \rho_0^{\Gamma}$,
with $\kappa = 0.2$, $\Gamma = 4/3$. This fluid is threaded by an
initially uniform magnetic field, everywhere directed along the $z$ axis (i.e. parallel to the
orbital angular momentum of the binary). Our canonical initial fluid density and magnetic field
strengths are $\rho_0 = 1, b_0 = 10^{-1}$ in code units; this is equivalent to
$B_0 = 3.363 \times 10^4 \, \UNITG$ for a physical density of $10^{-11} \, \UNITg \, \UNITcm^{-3}$,
or $B_0 = 3.363 \times 10^3 \, \UNITG$ for a physical density of $10^{-13} \, \UNITg \, \UNITcm^{-3}$.

\subsection{Diagnostics}

To better interpret the results of our simulations, we rely on several
diagnostics of the plasma and the black hole geometry. For
completeness, we describe these here.

To assess the extent of induced rotation for the system, we measure the fluid's angular
velocity $\Omfluid$ about the orbital axis, defined as
\beq
\Omfluid = \frac{x v^y - y v^x}{(x^2 + y^2)}. \label{eq:om_rot}
\eeq

For a test particle moving around a Kerr black hole of mass $M$ and spin parameter
$a = J/M$, the Keplerian angular frequency is (see, e.g. \cite{Bardeen:1972fi})
\beq
\Omega_{K} = \frac{1}{M} \left[ \left(\frac{r_{\rm KBL}}{M}\right)^{3/2} + \frac{a}{M} \right]^{-1}, \label{eq:om_KeplerKBL}
\eeq
where $r_{\rm KBL}$ is the areal radius
\footnote{When working in simulation coordinates we deduce the areal radius from the form of a curvature invariant evaluated on the equatorial plane
on the same time slice, as was done for the ``Lazarus'' procedure \cite{Baker:2001sf,Campanelli:2005ia}.}
of Kerr-Boyer-Lindquist coordinates.

Another angular frequency of interest is that of a zero-angular-momentum particle infalling from
infinity in a Kerr spacetime:
\beq
\Omega_{\rm infall} = \frac{2 M a}{(r^3_{\rm KBL} + a^2 r_{\rm KBL} + 2 M a^2) }. \label{eq:om_radinfall}
\eeq

The relativistic Alfv\'en velocity of the magnetized fluid is defined as \cite{Gedalin_1993}
\begin{align}
\valf &= \sqrt{\frac{b^2}{\rho (1 + \epsilon) + \press + b^2}} = \sqrt{\frac{b^2}{ \rho (1 + \Gamma \epsilon) + b^2}} \nonumber \\
      &= \sqrt{\frac{b^2}{\rho + 4 \press + b^2}},  \label{eq:valf_def}
\end{align}
where the second line holds for a polytrope with $\Gamma = 4/3$, as we use here.

To make contact with the results of \cite{Giacomazzo:2012iv}, we look primarily at the Poynting
vector. In terms of the MHD fields evolved, this can be calculated as
\beq
S^i \equiv \alpha T^i_{{\rm EM}, 0} = \alpha \left( b^2 u^i u_0 + \frac{1}{2} b^2 g^i_{\;0} - b^i b_0 \right). \label{eq:Poynting_def}
\eeq
We frequently use $\Szdom$, the $(l=1,m=0)$ spherical harmonic mode of $S^z$, as a measure
of Poynting luminosity:
\beq
\LEM \approx \lim_{R \rightarrow \infty} \oint R^2 S^z \cos\theta d\Omega = \lim_{R \rightarrow \infty} 2 R^2 \sqrt{\frac{\pi}{3}} \Szdom. \label{eq:LEM_as_Sz10}
\eeq
In Appendix~\ref{apx:LEM_Sz10}, we justify this choice, and relate it to the EM flux measured by
\cite{Neilsen:2010ax,Palenzuela:2010xn}.

To estimate the rate of accretion of fluid into the black holes, we use the \texttt{Outflow} code
module in the Einstein Toolkit. \texttt{Outflow} calculates the flux of
fluid across each apparent horizon $S$ via:
\beq
\dot M = -\oint_{S} \sqrt\gamma \alpha D \left( v^i - \frac{\beta^i}{\alpha} \right) d\sigma_i, \label{eq:Mdot_formula}
\eeq
where $D \equiv \rho \alpha u^0$ is the Lorentz-weighted fluid density, and $\sigma_i$ is the
ordinary (flat-space) directed surface element of the horizon. BH
apparent horizons are located using the \texttt{AHFinderDirect} code
\cite{Thornburg:2003sf}.

\subsection{Comparison with Whisky 2012 Results}
\label{ssec:Whisky2012comp}

In this paper we apply recent advances in numerical
relativity techniques encoded in \IGM{} to achieve
longer-duration simulations covering a broader variety of simulation
scenarios than those studied in~\cite{Giacomazzo:2012iv} using \Whisky. As a
preliminary step, we first make contact with those earlier results, treating
the same scenario with the new numerical methods.

While \IGM{} is a newer code than \Whisky{}, its lineage traces back
more than a decade to the development of the Illinois Numerical
Relativity group's original GRMHD code \cite{Duez:2006qe,Etienne:2010ui,Etienne:2011re}
The algorithms underlying \Whisky{} and \IGM{} were chosen
through years of trial and error to maximize robustness and
reliability in a variety of dynamical spacetime contexts:
the Piecewise Parabolic Method \cite{Colella:1982ee} for
reconstruction, the Harten-Lax-van Leer approximate Riemann solver,
and an AMR-compatible vector-potential formalism for both evolving the
GRMHD induction equation and maintaining divergenceless magnetic
fields. 

Despite their algorithmic similarities, \Whisky{} and \IGM{}
were developed independently and as such, adopted formalisms and
algorithmic implementations are different. Most of these differences
should largely result in solutions that converge with increasing grid
resolution. For example, \IGM{} reconstructs the 3-velocity that appears in the induction
equation, $v^i \equiv u^i/u^0$ and \Whisky{} chooses to reconstruct the ``Valencia'' 3-velocity
$v_{(n)}^i \equiv (u^i/u^0 + \beta^i)/\alpha$.
Also, \Whisky{} defines the vector potential at vertices on our Cartesian grid, while \IGM{}
adopts a staggered formalism~\cite{BalsaraSpicer_1999}.

Beyond algorithmic implementations, two key choices made in the 2012
\Whisky{} paper \cite{Giacomazzo:2012iv} may result in significant
differences with this work. First, in \cite{Giacomazzo:2012iv}, \Whisky{}
actively maintained the exact polytropic relationship 
\eqref{eq:polytrope_EOS}, while with the new \IGM{} evolutions, the value of
$\kappa$ is allowed to change. This means that in the principal
simulations of \cite{Giacomazzo:2012iv}, no shock heating was allowed.

\begin{figure}
  \includegraphics[trim=5mm 0mm 0mm 0mm,clip,scale=0.60]{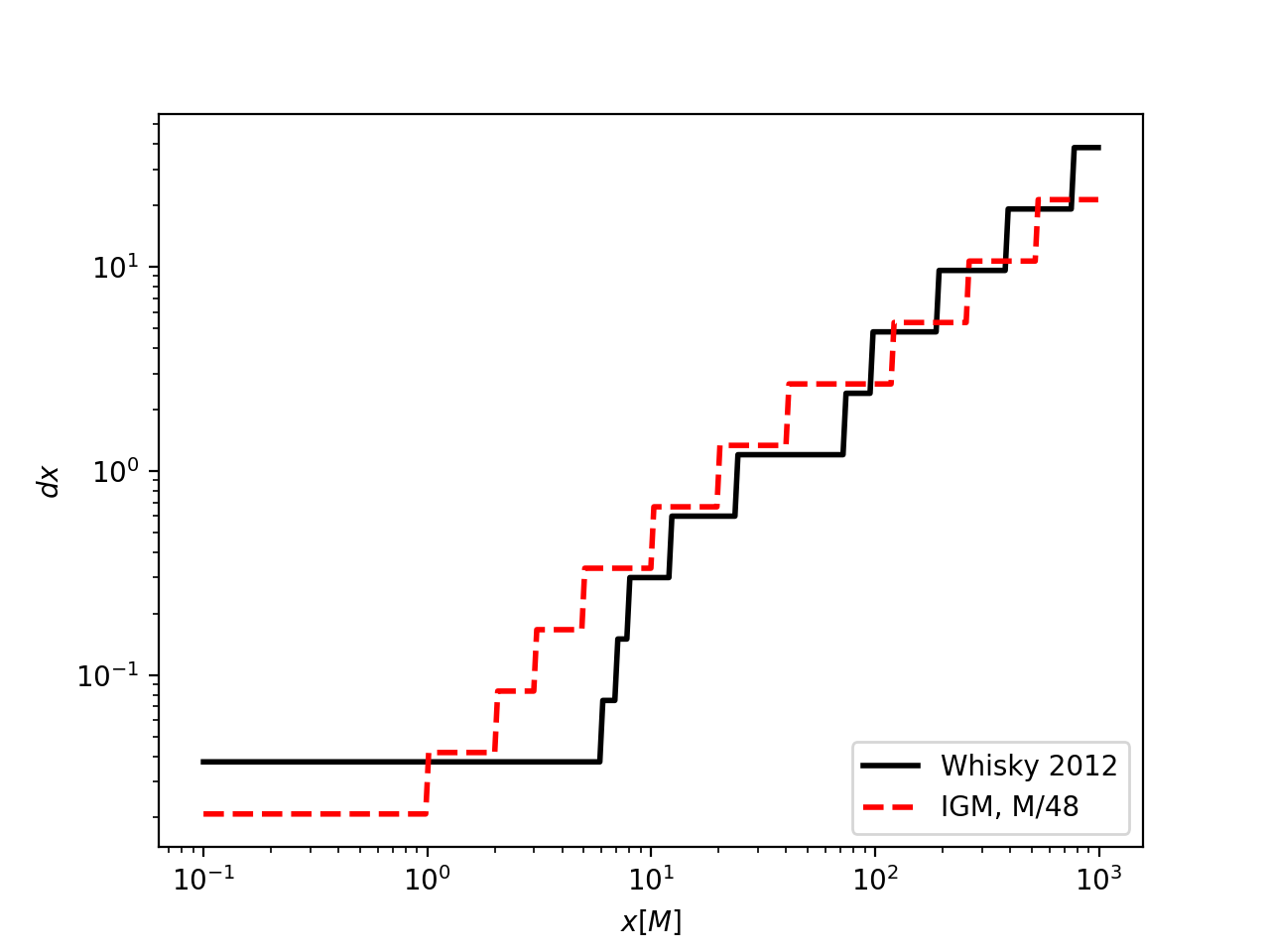}
  \caption{Initial numerical mesh refinement structure for original \Whisky{} run of \cite{Giacomazzo:2012iv} (black, solid),
  and lowest-resolution \IGM{} runs (red, dashed), expressed as local grid spacing $dx$ as a
  function of ``radial distance'' $x$ from the puncture at the center of the grid. Both grids have 11 refinement levels, with
  similar outer resolutions. The \IGM{} run is better-resolved in the regions immediately around the individual horizons, but the
  \Whisky{} run maintained uniformly high resolution for all points within $6M$ of each puncture.}
  \label{fig:Whisky-IGM_grid_comparison}
\end{figure}

Second, the electromagnetic gauge condition adopted in \cite{Giacomazzo:2012iv}
was later found to exhibit zero-speed modes that manifest as an
accumulation of errors at AMR grid boundaries \cite{Etienne:2011re}.
The impact of these gauge modes was somewhat mitigated
by the choice of very large high-resolution AMR grids near
the binary. \IGM{} adopts a generalization of the Lorenz
gauge \cite{Farris:2012ux} that removes the
zero-speed modes, and thus enables us to select a more optimal AMR grid
structure for the problem. 
To this end, Fig.~\ref{fig:Whisky-IGM_grid_comparison} presents the initial set of refinement
``radii'' (actually cube half-side) and associated resolutions for both the \Whisky{} and the
standard low-resolution \IGM{} runs.
While the \Whisky{} runs have a higher resolution throughout the wider region of radius
$1 M \lesssim r \lesssim 7 M$ centered on each puncture, the new \IGM{} runs better resolve the
region immediately around ($r \lesssim 1 M$) each black hole. The lower \Whisky{} resolution around
the horizons had a significant impact on the BH dynamics: with the
grids used in the original \Whisky{} runs, the black holes
merge at $t_{\rm merge} \sim 350M$, compared with $t_{\rm merge} \sim
450M$ for grids used in the \IGM{} runs presented here.

\begin{figure}
  \includegraphics[trim=5mm 0mm 0mm 0mm,clip,scale=0.60]{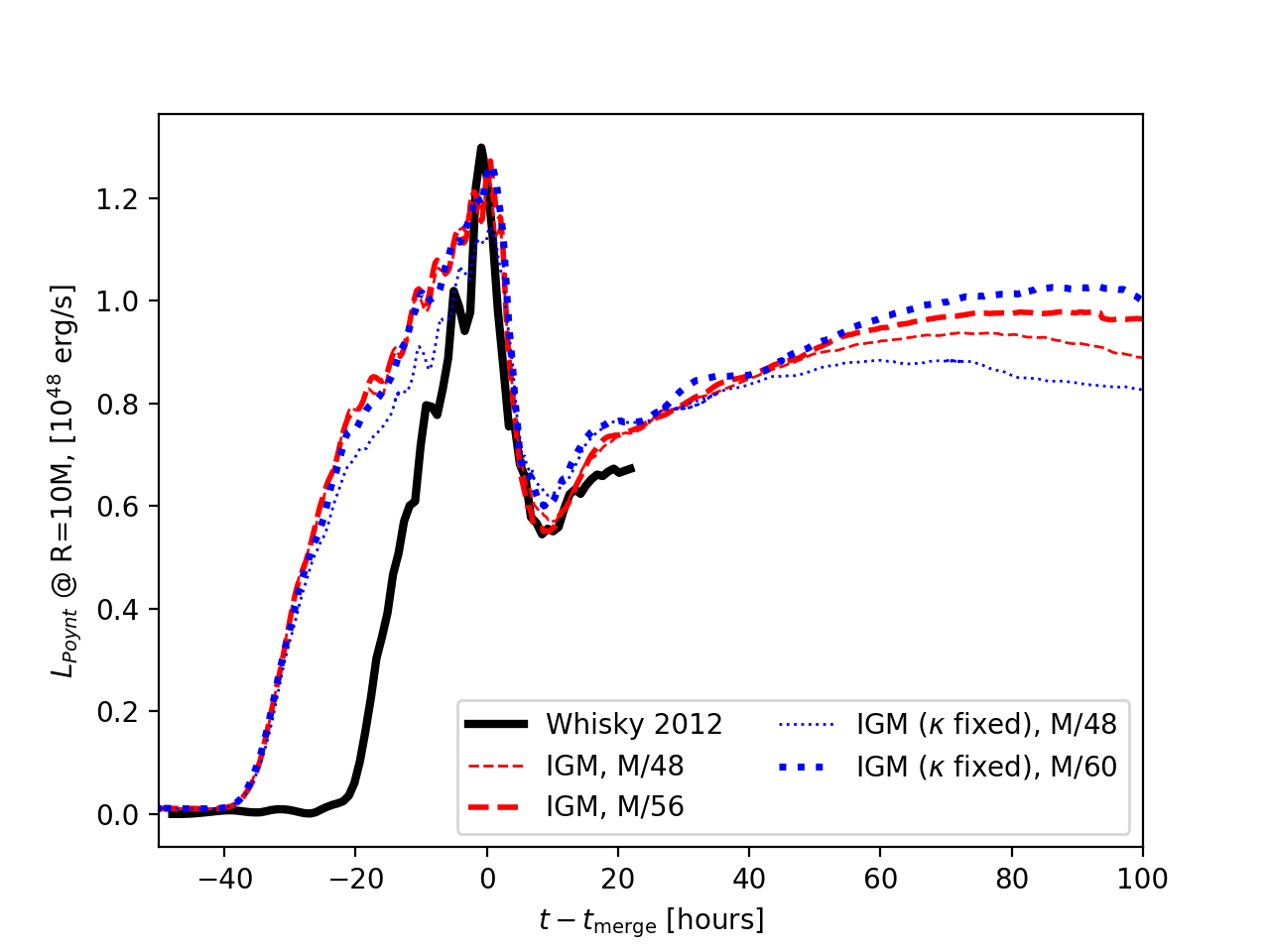}
  \caption{$\LEM$ for original \Whisky{} run of \cite{Giacomazzo:2012iv} (black, solid), compared with the new
  \IGM{} runs for the same initial separation (red, dashed). All luminosities have been time-shifted by the time of
  merger for that run, and scaled to reflect the canonical case in \cite{Giacomazzo:2012iv}: a plasma of uniform initial density
  $\rho_0 = 10^{-11} \, \UNITg \, \UNITcm^{-3}$ and magnetic field strength $B_0 = 3.363 \times 10^4 \, \UNITG$, in the vicinity of a
  black-hole binary of total mass $\MTOT = 10^8 \MSun$. An \IGM{} simulation keeping the polytropic coefficient $\kappa$
  fixed to its initial value everywhere (i.e., disabling shock heating) shows very similar behavior (blue, dotted).}
  \label{fig:LEM_Whisky2012_comparison_CGS}
\end{figure}

In Fig.~\ref{fig:LEM_Whisky2012_comparison_CGS}, we show the resulting Poynting luminosity
from both the \Whisky{} run and the new \IGM{} run.
\footnote{Note that Fig.~5 from \cite{Giacomazzo:2012iv} computes the
  luminosity only for $z>0$; we multiply the 2012 result by two here
  to compensate.}
The peak luminosity is very similar in both cases,
but the rise to this peak is sharper in the \Whisky{} case because
under-resolved horizon regions result in a considerably earlier
merger time of the black holes in the \Whisky{} run.
We have verified that the different treatment of the polytropic coefficient $\kappa$ between
\Whisky{} and \IGM{} has minimal effect on the luminosity, by performing a modified \IGM{} simulation
with fixed $\kappa$ (i.e., with shock-heating disabled) (blue curve
in Fig.~\ref{fig:LEM_Whisky2012_comparison_CGS}).

\section{Results}
\label{sec:results}

Our simulations are designed to explore the MHD physics that may give
rise to EM counterparts to black hole mergers. These simulations,
however, are not appropriate over the large temporal and spatial
scales required to simulate the emission of EM radiation to a very
distant observer (``at infinity''); the black hole region is fully
enshrouded by an infinite region of finite-density gas which would
soon block any radiation or other outflows. Our focus instead is to
examine near-zone mechanisms that could drive EM outflows. Two broad
channels of emission are considered. First, the development of
familiar jet-like structures leading to strong Poynting flux on the
axis can provide a significant source of energy, which can be
converted to strong EM emissions farther downstream. Second, we also
consider mechanisms for direct emission from the fluid flows near the
black holes, ignoring the absorbing properties of matter farther out.

Our canonical configuration is an equal-mass BBH with initial
coordinate separation $d = 14.4 M$, initial fluid density $\rho_0 = 1$
in a polytrope with $\kappa = 0.2, \Gamma = 4/3$, and initial magnetic
field strength $b_0 = 0.1$. We present these and derived parameters in
Table~\ref{tab:canonical_params}.
 
\begin{table}\footnotesize
\centering
\caption{Initial parameters and derived quantities for the canonical configuration: initial
         puncture separation $d$, puncture mass $m_{\rm p}$, Bowen-York linear momentum components
         $P_{\rm tang}$ \& $P_{\rm rad}$, finest grid spacing $dx$, merger time $\tmerge$,
         initial fluid density $\rho_0$, magnetic field strength $b_0$, polytropic constant
         $\kappa_0$, fluid pressure $\press_0$, specific internal energy $\epsilon_0$, ratio of magnetic to fluid energy
         density $\umagofluid_0$, specific enthalpy $h^*_0$, and ambient Alfv\'en speed $\valf$.}
\begin{tabular}{rrrrr}
\hline \hline
$d (M)$ & $m_{\rm p}$   & $P_{\rm tang} (M)$ & $P_{\rm rad} (M)$ & $dx (M)$ \\
\hline
14.384  & 0.4902240 &  0.07563734  & -0.0002963 & 1/48\\
\hline \hline
\end{tabular}
\begin{tabular}{r|rrr|rrrrr}
 $\tmerge (M)$ & $\rho_0$ & $b_0$ & $\kappa_0$ & $\press_0$ & $\epsilon_0$ &  $\umagofluid_0$ & $h^*_0$ & $\valf$ \\
\hline
 3514.333          & 1.0      & 0.1   & 0.2        & 0.2   &  0.6         & 5.0e-3  & 1.81 & 0.07433 \\
\hline \hline
\end{tabular}
\label{tab:canonical_params}
\end{table}

\subsection{Large-Scale Structure of Fluid and Fields}
\label{ssec:results_global}

We begin by presenting an overview of the major field structures that develop through MHD dynamics
during the merger process, using our canonical case as a representative example.

\begin{figure}
  \includegraphics[trim=0mm 0mm 0mm 0mm,clip,width=0.5\textwidth]{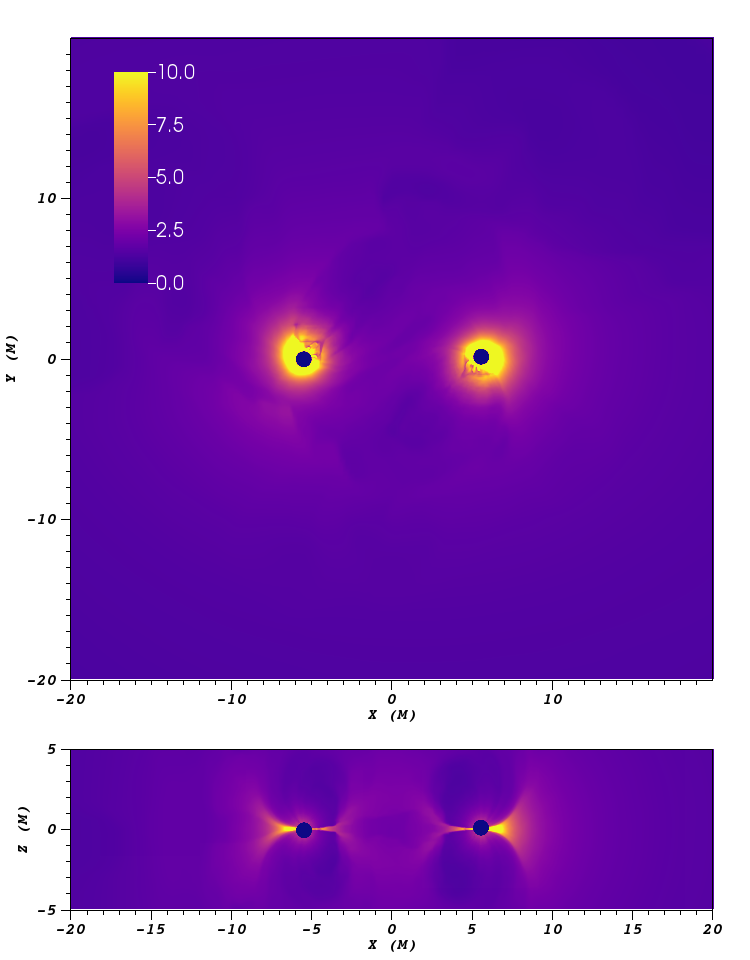}
  \caption{Fluid density $\rho$ during inspiral at time $t = 2400M$ (about $1100M$ before merger) for the $d = 14.4M$ configuration.
           At this time the holes are centered at $(x, y) \approx \pm (5.53M, 0.08M)$. 
           The regions inside the BH horizons have been masked
           out. Note that in all configurations
           the BHs are orbiting in a counter-clockwise motion around the positive $z$ axis}
  \label{fig:rho_d14p4_t2400}
\end{figure}

The canonical simulation begins about $3500M$ before merger, with an initially uniform
fluid and a uniform vertical magnetic field. After some time the fluid has fallen mostly
vertically along the field lines, concentrating in a nearly
axisymmetric thin disk ($h \ll M$) of dense material about each black hole. 
Figure~\ref{fig:rho_d14p4_t2400} shows a snapshot of the fluid density $\rho$ on the $x$-$y$
(orbital) and $x$-$z$ planes during the late inspiral (about $1100M$ before merger) for the
$d = 14.4M$ configuration.

\begin{figure}
  \includegraphics[trim=0mm 0mm 0mm 0mm,clip,width=0.5\textwidth]{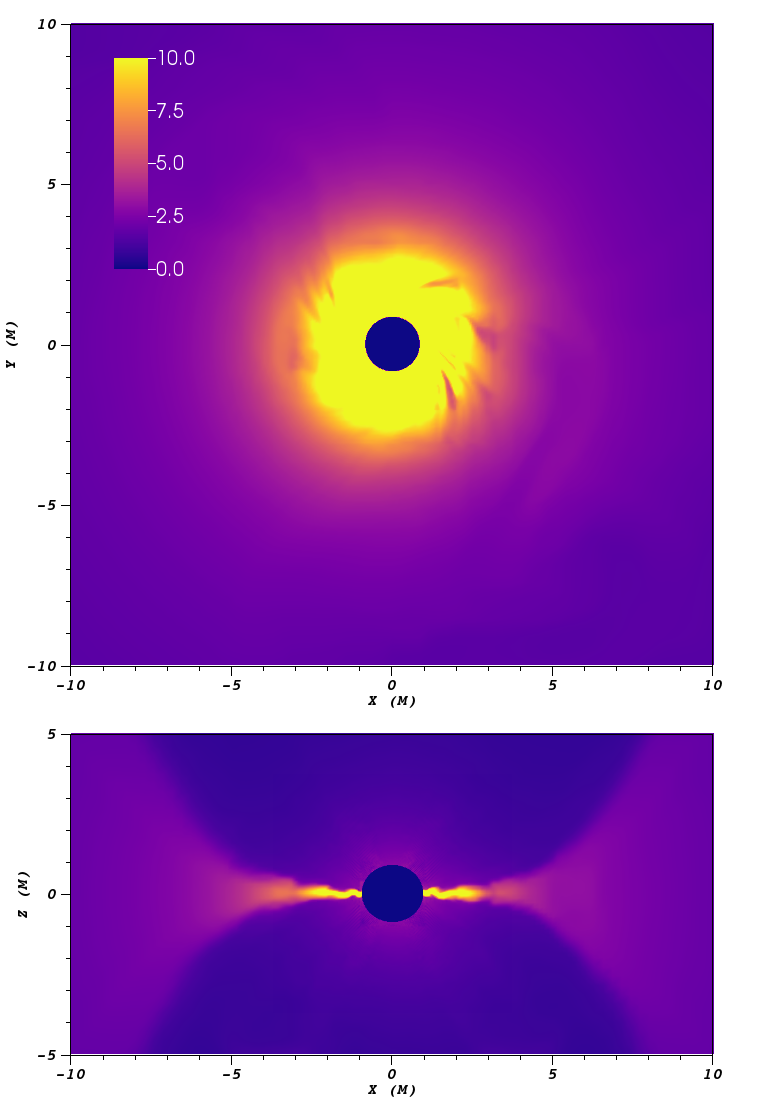}
  \caption{Fluid density $\rho$ approximately $1500 M$ after merger
    for the $d = 14.4M$
    configuration. The region inside the BH horizon has been masked out.}
  \label{fig:rho_d14p4_t5000}
\end{figure}

By late times, those disks have merged into a common disk around the
final, spinning black hole. The structure of the post-merger disk is shown
in Fig.~\ref{fig:rho_d14p4_t5000}, where we again plot $\rho$ on the
$x$-$y$ and $x$-$z$ planes. By this time fluid has fallen in to form a thin disk ($h \ll M$) of dense material
with radius of $2-3$ gravitational radii (the BH horizon radius is approximately $1M$ here). Above
and below the disk, gas is largely excluded by magnetically dominated regions. Focusing just on the
$x$-$y$ plane, the top panel shows that some asymmetric structure persists long after merger.

\begin{figure}
  \includegraphics[trim=0mm 0mm 0mm 0mm,clip,width=0.5\textwidth]{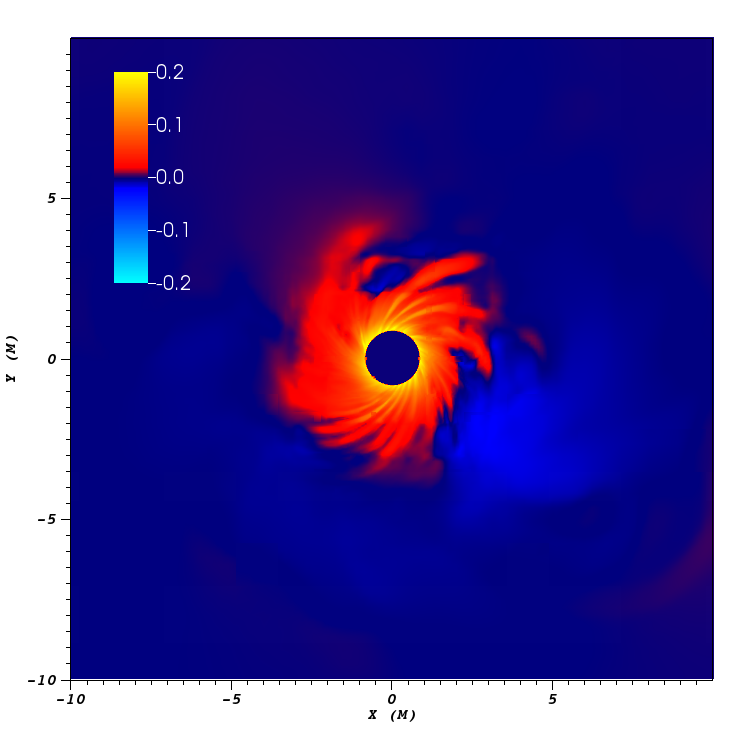}
  \includegraphics[trim=5mm 0mm 0mm 0mm,clip,width=0.5\textwidth]{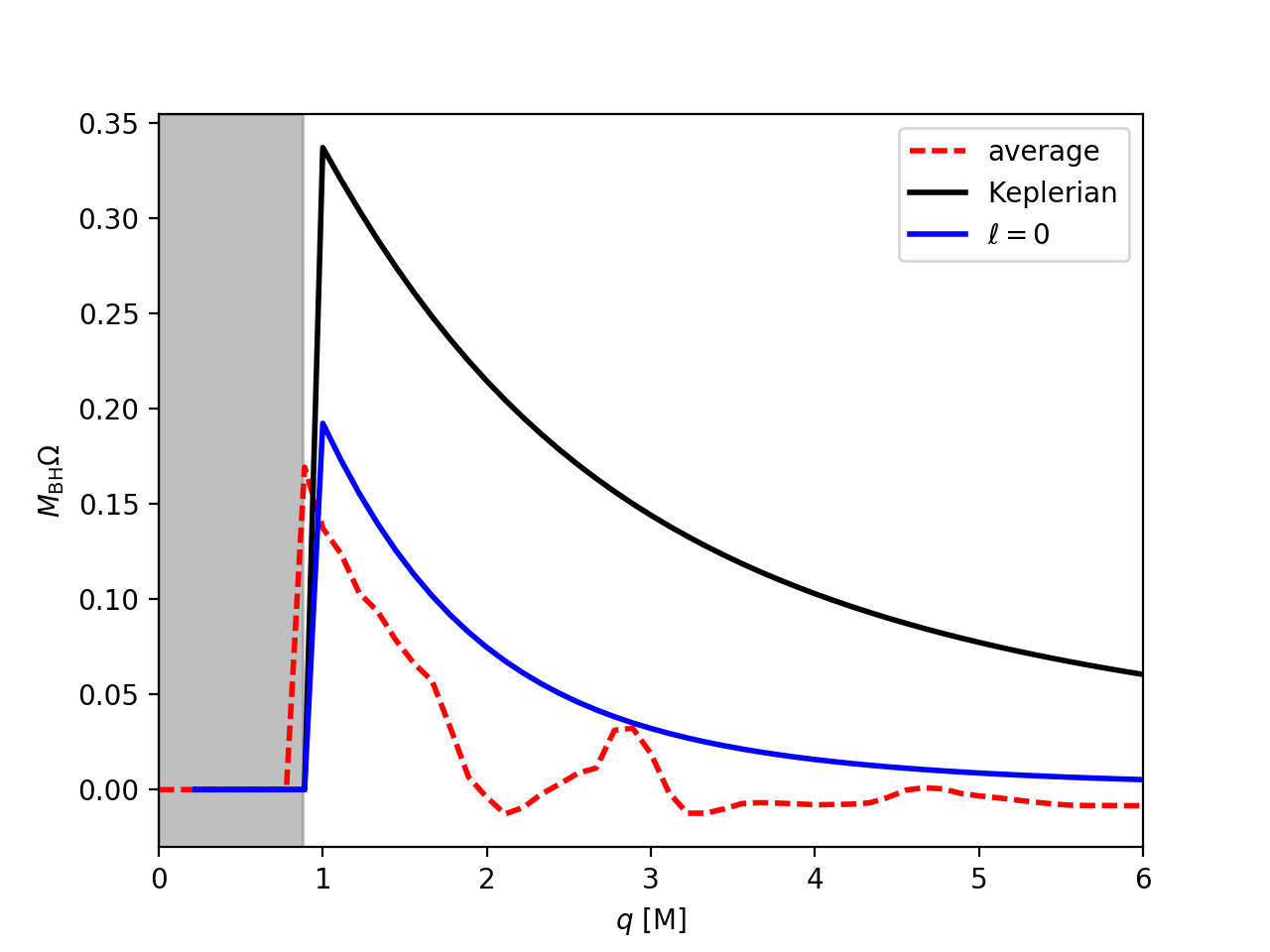}
  \caption{{\bf Top panel}: Post-merger fluid orbital frequency $\Omfluid$ for the high-resolution $d = 14.4M$ configuration.
  The black-hole interior has been masked out. {\bf Bottom panel}: $M_{\rm BH} \Omfluid$ as function of cylindrical radius $q$, averaged over
  orbital azimuthal angle (red dashed), with the corresponding relativistic Keplerian angular frequency $M_{\rm BH} \Omega_{K}$
  (black solid), and the angular frequency induced for a zero-angular-momentum ($\ell = 0$) infalling test particle
  (blue solid). The shaded region marks the interior of the black-hole horizon.}
  \label{fig:Omega_XY_t4900M_d14p4}
\end{figure}

Though these pre-merger and post-merger disks superficially resemble familiar black hole accretion disks,
there are important differences. Traditional disks are centrifugally
supported outside the innermost stable circular orbit. Our fluid
distribution, on the other hand, is initially at
(coordinate) rest with low specific angular momentum. While these flows are stirred first by binary
motion and later by frame-dragging near the final spinning black hole, this does not produce
a Keplerian flow. This can be seen in Fig.~\ref{fig:Omega_XY_t4900M_d14p4}, which shows 
the fluid orbital frequency $\Omfluid$ \eqref{eq:om_rot} about $1100M$ after merger.
The region around the horizon exhibits a spin-up of the fluid material for
$r \lesssim 4M$ to an angular frequency of up to $M \Omfluid \sim 0.1$. This
can be compared with two other angular velocity profiles of interest:
the Keplerian angular velocity $\Omega_{K}$ \eqref{eq:om_KeplerKBL}
for a rotationally supported disk, and the ``infall angular velocity''
\eqref{eq:om_radinfall} for equatorial infall geodesics with vanishing
specific angular momentum.  Each is evaluated for the same Kerr BH 
($a = 0.69 M$). The velocity profile of our disk more closely
resembles the profile of infalling geodesics.

\begin{figure}
  \includegraphics[trim=0mm 0mm 0mm 0mm,clip,scale=0.5]{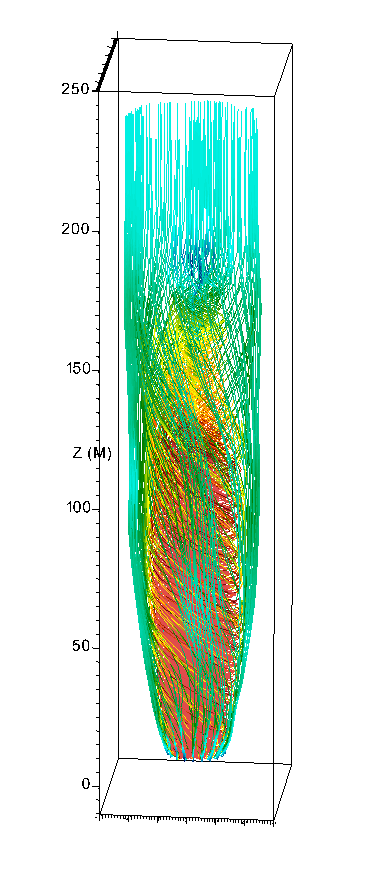}
  \caption{Magnetic field streamlines in the polar region, around $1100M$ after merger. 
    The field lines are twisted into a helical pattern, concentrated at
    the origin. This helical structure propagates outward at the ambient Alfv\'en speed $\valf = 0.07433$,
    replacing the initially vertical $\ve{B}$ fields (still visible at large $z$).}
  \label{fig:Bvec_streamlines_d11p5_t2500}
\end{figure}

During evolution, the initially parallel, $z$-directed magnetic field lines evolve to resemble the
structure of a black-hole jet. The field lines are pinched in the
orbital plane as the matter falls in through the disk region, and
become twisted into a helical structure---see
Fig.~\ref{fig:Bvec_streamlines_d11p5_t2500}---through the rotational
motion in the orbital/infall plane.
This structure originates in the strong-gravitational-field region and
propagates outward at the ambient Alfv\'en speed $\valf$ \eqref{eq:valf_def}.

\begin{figure}
  \includegraphics[trim=5mm 0mm 0mm 0mm,clip,width=0.4\textwidth]{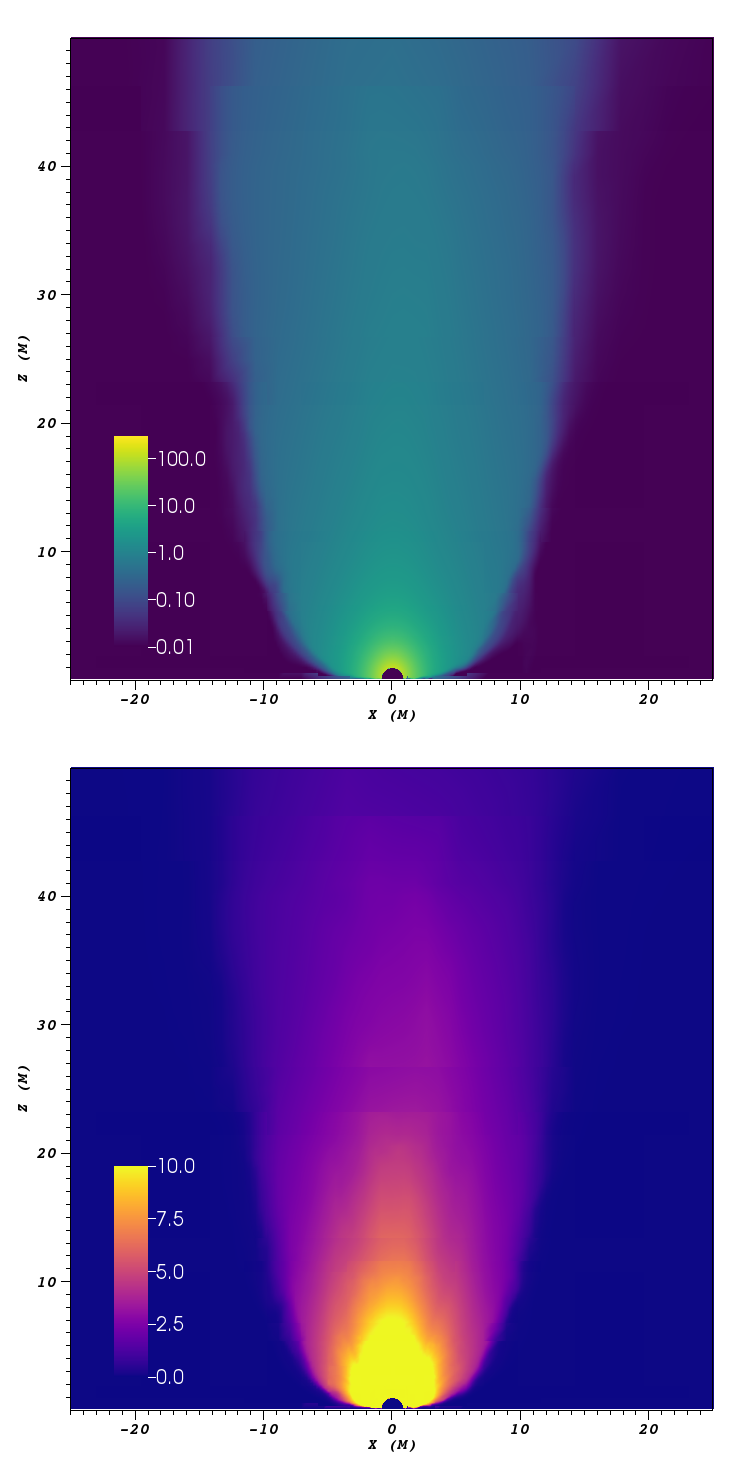}
  \caption{{\bf Top panel}: Magnetic field squared magnitude $b^2$ about $1100M$ after merger
           for the high-resolution $d = 14.4M$ configuration.
           {\bf Bottom panel}: Magnetic-to-gas pressure ratio $\beta^{-1} \equiv b^2/2 \press_{\rm gas}$ for
           the same time and configuration.}
  \label{fig:smallb2_betainv_X14p4_t4931M}
\end{figure}

\begin{figure}
  \includegraphics[trim=5mm 0mm 0mm 0mm,clip,width=0.4\textwidth]{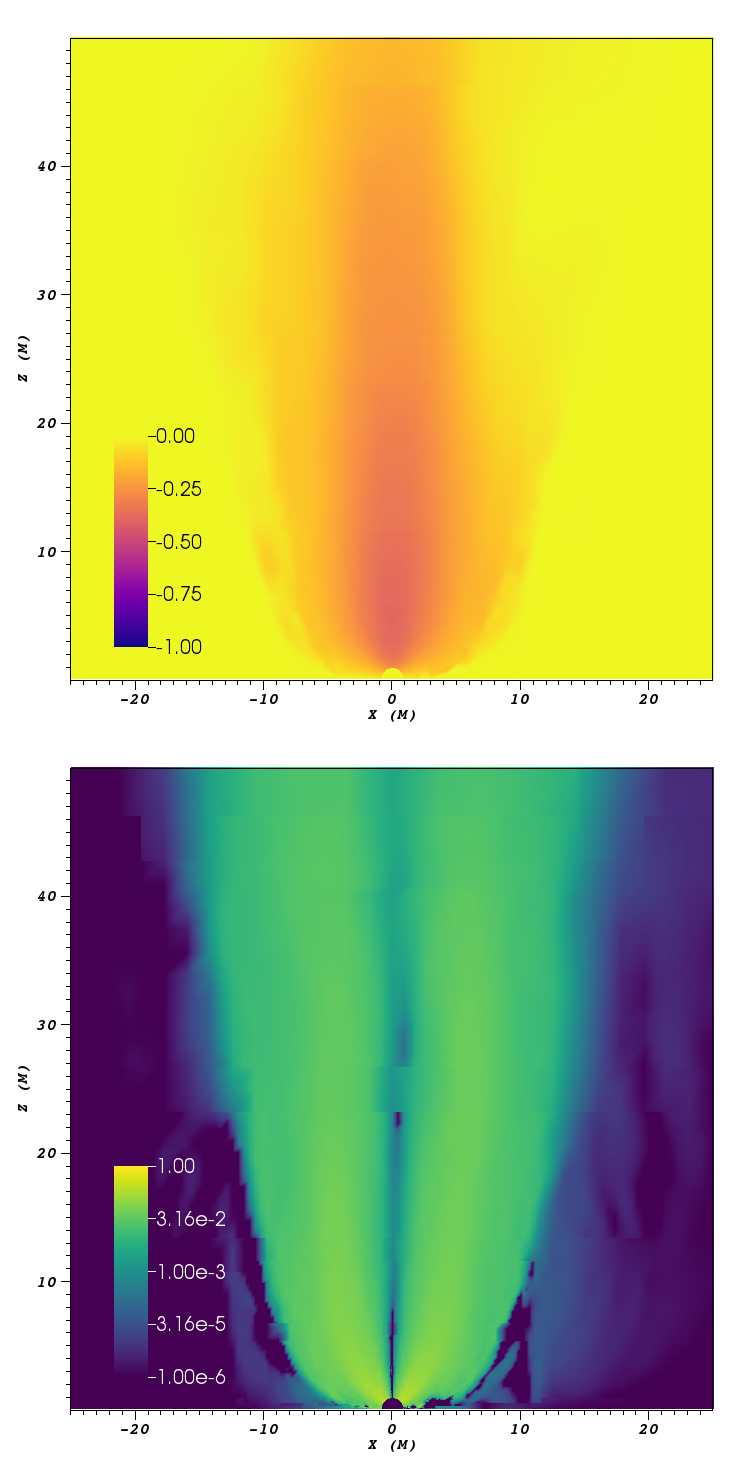}
  \caption{{\bf Top panel}: Fluid velocity ($z$ component) about $1100M$ after merger
           for the high-resolution $d = 14.4M$ configuration.
           {\bf Bottom panel}: Poynting vector \eqref{eq:Poynting_def} ($z$ component) for the same time and configuration.}
  \label{fig:vz_Poynz_X14p4_t4931M}
\end{figure}

This process also enhances the magnetic field strength in the region above and below the orbital plane.
In Fig.~\ref{fig:smallb2_betainv_X14p4_t4931M}, we show the state of the evolved (squared) magnetic field
strength $b^2$ $1100M$ after merger,
evaluated on the $x$-$z$ plane.
As seen in the top panel, $b^2$ is greatly amplified at and near the polar axis of the post-merger
hole. The lower panel shows that this region is dominated by magnetic pressure. 
This region shares some features of a relativistic jet,
as both are magnetically dominated and contain a helical magnetic field
structure. We show in Fig.~\ref{fig:vz_Poynz_X14p4_t4931M} that the
structures we observe yield a strong Poynting flux directed
\emph{outward}. As with our disk however, through the course of these
simulations the fluid flow through these jet-like structures is
predominantly \emph{inward}-directed. Nonetheless, over longer temporal
and larger spatial scales and in plausible astrophysical environments,
the strong Poynting flux could drive relativistic outflows and strong EM
emissions.  We further explore this as a source of energy to
eventually power EM counterparts in the next section.\footnote{ 
  There is no direct contradiction between inward fluid flows and outward Poynting flux.  A simple expression relating Poynting flux to velocity is
  $\LEM^z=B^2v^z_\perp$, where $v^z_\perp=v^z-v^z_\parallel$ is the component of fluid velocity perpendicular to the magnetic field lines.  For a specified
  Poynting flux, the parallel component of velocity $v^z_\parallel$ is not directly constrained and may be negatively directed and large enough to overcome a positive
  $v^z_\perp$.}

\subsection{Mass Accretion Rate}
\label{ssec:results_mdot}

\begin{figure}[h]
\includegraphics[width=0.5\textwidth]{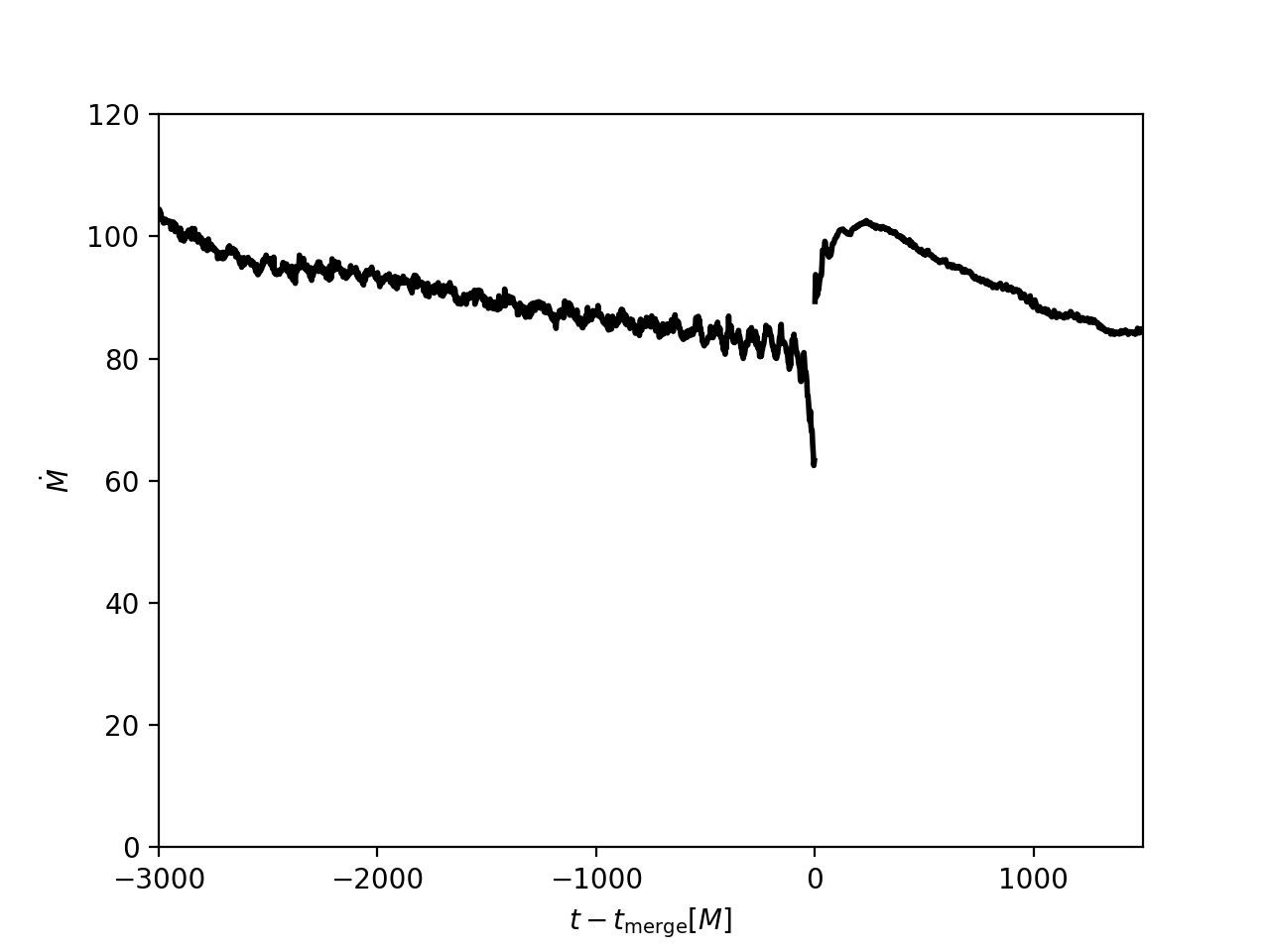}
\caption{Rate of mass loss $\Mdot$ to accretion into the black hole horizons.}
\label{fig:MdotFromOutflow}
\end{figure}

Although the initially static fluid in our simulations does not
develop the rotational support necessary for an accretion disk (as
Fig.~\ref{fig:Omega_XY_t4900M_d14p4} indicates), the rate of
accretion $\Mdot$ onto the black holes provides a measure of the
energy available for EM outflows during inspiral and merger. In
Fig.~\ref{fig:MdotFromOutflow}, we show the development of this
quantity over the bulk of the $d = 14.4M$ evolution, calculated using
\eqref{eq:Mdot_formula}.

We note the main features of this accretion rate estimate: (i) $\Mdot$
slowly declines through the late inspiral, with the
drop-off steeper just before merger when a common horizon forms; (ii)
$\Mdot$ jumps when the black hole apparent horizons join
discontinuously at merger; (iii) after some settling in, the
post-merger $\Mdot$ resumes the slow decline seen before merger.

The numbers in Fig.~\ref{fig:MdotFromOutflow} are in code units where
$\MTOT = 1$, $\rho_0 = 1$. Since $\Mdot$ generically scales as
$\rho\MTOT^2$, we convert to physical units using a factor
$G^2/c^3$. Scaling for our canonical initial fluid density and system
mass, we obtain the rate in cgs units as
\beq
\Mdot_{\rm cgs} = 6.54 \times 10^{23} \rhothir M_{8}^2 \Mdot \, \UNITg \, \UNITs^{-1},
\eeq
where $\rhothir \equiv \rho_0/(10^{-13} \, \UNITg \, \UNITcm^{-3})$,
and $M_8 \equiv \MTOT/(10^8 \MSun)$.

Since $\Mdot \sim 100$ throughout the simulation, a good order-of-magnitude estimate
for the accretion rate both before and after merger is
$\Mdot_{\rm cgs} \approx 6 \times 10^{25} \rhothir M_{8}^2 \, \UNITg \, \UNITs^{-1}$.

\subsection{Features of Poynting Luminosity}
\label{ssec:results_poynting}

The powerful Poynting flux generated by our simulations shows that strong flows of
electromagnetic energy are driven vertically outward along the orbital
angular momentum axis, starting near the orbital plane.
Many studies have shown that such Poynting flux regions can transfer
power from the black hole region, driving relativistic outflows
\cite{Blandford:1977ds,Paschalidis:2014qra,Ruiz2016}, and then through
a cascade of internal or external matter interactions, ultimately
yielding strong EM emissions (e.g., in the fireball model for
gamma-ray bursts \cite{Piran:1999kx}). Our simulations are not
set up to model those processes, but we can explore the Poynting
luminosity as a \emph{potential} source of power for EM counterpart
signals.

To get a measure of time dependence of the jet-like Poynting-driven EM power, we compute the
Poynting luminosity $\LEM$ from \eqref{eq:LEM_as_Sz10}, using the dominant $(l,m) = (1,0)$
spherical harmonic mode of the $z$-component of the Poynting flux, $S^z$ \eqref{eq:Poynting_def},
extracted on a coordinate sphere of radius $R = 30 M$. Results from
this diagnostic are shown in Fig.~\ref{fig:LEM_annotated}. As
discussed in Appendix~\ref{apx:LEM_Sz10}, this rotation-axis-aligned component dominates the
Poynting flux: $S^r \approx S^z \cos\theta$.
We select extraction at $30M$ as giving a measure of the input energy for potential reprocessing into
EM signals down stream. This extraction radius is far enough to avoid confusion with the motion of
the black holes, yet close enough to provide a quick measure of potential emission on timescales
comparable to the merger-time.
\footnote{In \cite{Giacomazzo:2012iv}, extraction was carried out at $R = 10M$, but the initial
binary separation was much smaller in that case.}

Several features are evident in Fig.~\ref{fig:LEM_annotated}: (a) an early local maximum in the
flux (occurring at $t \sim 100 M$ for this extraction radius);  (b) a steep rise in flux amplitude
beginning at $t \sim 450 M$, followed by (c) a slight drop to a slow-growth stage, ending in a
rapid climb and with a slight ``blip'' (d), leading to a final maximum value (e) before a gradual
fall-off. We believe
that these features correspond to (a) the initial settling of the
GRMHD fluids and black hole space-time, (b) the arrival of
magnetic-field information from the black hole region at the
extraction sphere, (c) development relating to the inspiral process,
(d) prompt response to merger, and (e) initiation of single-black
hole jet-like characteristics.

\begin{figure}
  \includegraphics[trim=0mm 0mm 0mm 0mm,scale=.60]{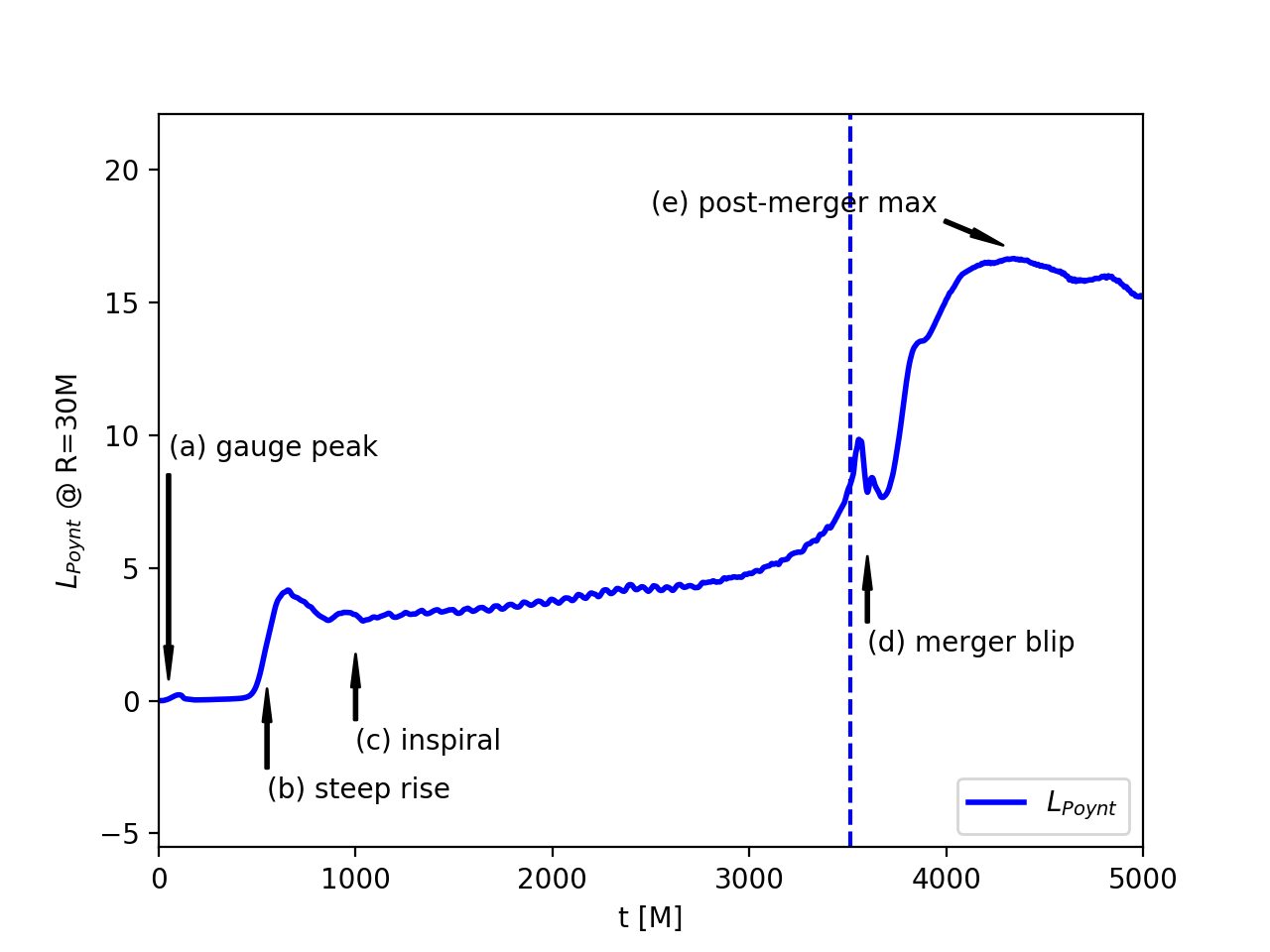}
  \caption{$\LEM$, the Poynting luminosity, for the $d = 14.4M$ configuration considered in
           Table~\ref{tab:merge_info}; extraction of the mode is on a coordinate sphere of radius
           $30 M$. The merger time is marked by a dashed vertical line.}
  \label{fig:LEM_annotated}
\end{figure}

\subsubsection{Dependence on Initial Separation}
\label{sssec:separations}

The plasma in our simulations is initially at rest near the black
holes, which is clearly unphysical. We must therefore be careful
to start our BBH at a large enough separation so that plasma in the
strong-field region has time to establish a quasi-equilibrium flow with the
binary motion.

Binary parameters for simulations covering a range of initial
separations are presented in Table~\ref{tab:ID_params}. To treat the
limit of zero initial 
separation, we also performed a simulation of a single Kerr black hole
(using the quasi-isotropic form of exact Kerr \cite{Brandt:1996si})
with parameters chosen consistent with the end-state black hole
observed after merger: $m_{\rm Kerr} = 0.97 M$, $a/m_{\rm Kerr} = 0.69$.
 
\begin{table}\footnotesize
\centering
\caption{Bowen-York parameters of the numerical configurations used. The holes are non-spinning,
         and are initially separated in the $x$ direction. Our canonical configuration is
         shown in bold face.}
\begin{tabular}{l rrrr}
\hline \hline
run name                & $d (M)$ & $m_{\rm p}$   & $P_{\rm tang} (M)$ & $P_{\rm rad} (M)$\\
\hline
\texttt{X1\_d16.3}      & 16.267  & 0.4913574 &  0.07002189  & -0.0002001 \\
\texttt{X1\_d14.4}      & {\bf 14.384}  & {\bf 0.4902240} &  {\bf 0.07563734}  & {\bf -0.0002963} \\
\texttt{X1\_d11.5}      & 11.512  & 0.4877778 &  0.08740332  & -0.0006127 \\
\texttt{X1\_d10.4}      & 10.434  & 0.4785587 &  0.0933638   & -0.00085   \\
\texttt{X1\_d9.5}       &  9.46   & 0.4851295 &  0.099561    & -0.001167  \\
\texttt{X1\_d8.4}       &  8.48   & 0.483383  &  0.107823    & -0.0017175 \\
\texttt{X1\_d6.6}       &  6.61   & 0.4785587 &  0.1311875   & -0.0052388 \\
\hline
\hline \hline
\end{tabular}
\label{tab:ID_params}
\end{table}

\begin{table}\footnotesize
\centering
\caption{Time of merger $\tmerge$ for each binary configuration. As time of
  merger depends on resolution, we include resolution information for
  each case. Our canonical configuration is shown in bold face.}
\begin{tabular}{l rrrr}
\hline \hline
run name                &  $dx (M)$ & $\tmerge (M)$ \\
\hline
\texttt{X1\_d16.3}      & 1/48 & 5380 \\
\texttt{X1\_d14.4}      & {\bf 1/48} & {\bf 3514} \\
                        & 1/56 & 3651 \\
                        & 1/72 & 3797 \\
\texttt{X1\_d11.5}      & 1/48 & 1549 \\
                        & 1/56 & 1584 \\
                        & 1/72 & 1572 \\
\texttt{X1\_d10.4}      & 1/48 & 1054 \\
                        & 1/72 & 1066 \\
\texttt{X1\_d9.5}       & 1/48 & 681 \\
\texttt{X1\_d8.4}       & 1/48 & 451 \\
                        & 1/56 & 451 \\
\texttt{X1\_d6.6}       & 1/48 & 208 \\
\hline
\hline \hline
\end{tabular}
\label{tab:merge_info}
\end{table}

\begin{figure}
  \includegraphics[trim=5mm 5mm 5mm 5mm,scale=.60]{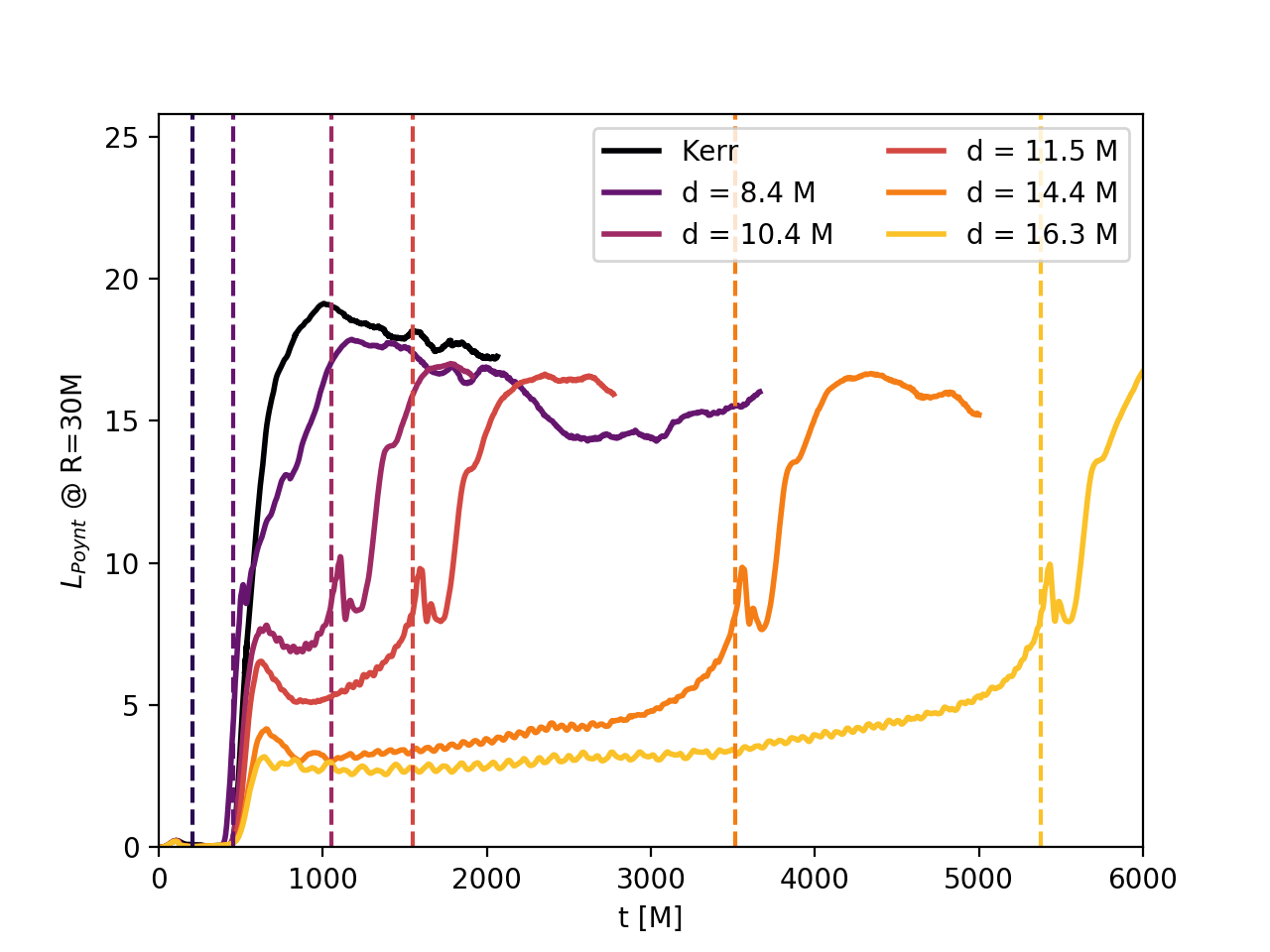}
  \caption{$\LEM$ for the configurations considered in Table~\ref{tab:merge_info}; extraction of the
  mode is on a coordinate sphere of radius $30 M$ for each case. Merger times for each binary
  are marked by dashed vertical lines. 
  ($1 \LEM = 5.867 \times 10^{44} \rhothir M_8^2 \, \UNITerg \, \UNITs^{-1}$.)
 }
  \label{fig:LEM_seps}
\end{figure}

In Fig.~\ref{fig:LEM_seps} we again show $\LEM$ at $R = 30 M$, but for
simulations beginning at times ranging from about $200M$ to $5400M$
before merger. For convenience, we show the merger time of each
configuration as a dashed line of the same color. While we generally
see the same set of features for each simulation, the time delay
between features (b) and (d) shrinks as the inspiral duration becomes
shorter.
The timing of features (a) and (b) indicates that they
can have no dependence on the merger of the binary, in contrast to the
conclusion drawn from the 2012 work~\cite{Giacomazzo:2012iv}. For
initially smaller-separation simulations such as the $d = 8.4M$ of
\cite{Giacomazzo:2012iv}, these features are poorly resolved; in
particular the ``slow-growth'' stage 
is almost completely absent. Consequently \cite{Giacomazzo:2012iv}
failed to distinguish the initialization-dependent rise (b) from the
inspiral- and merger-driven rise (c-e).

\begin{figure}
  \includegraphics[trim=5mm 5mm 5mm 5mm,scale=.60]{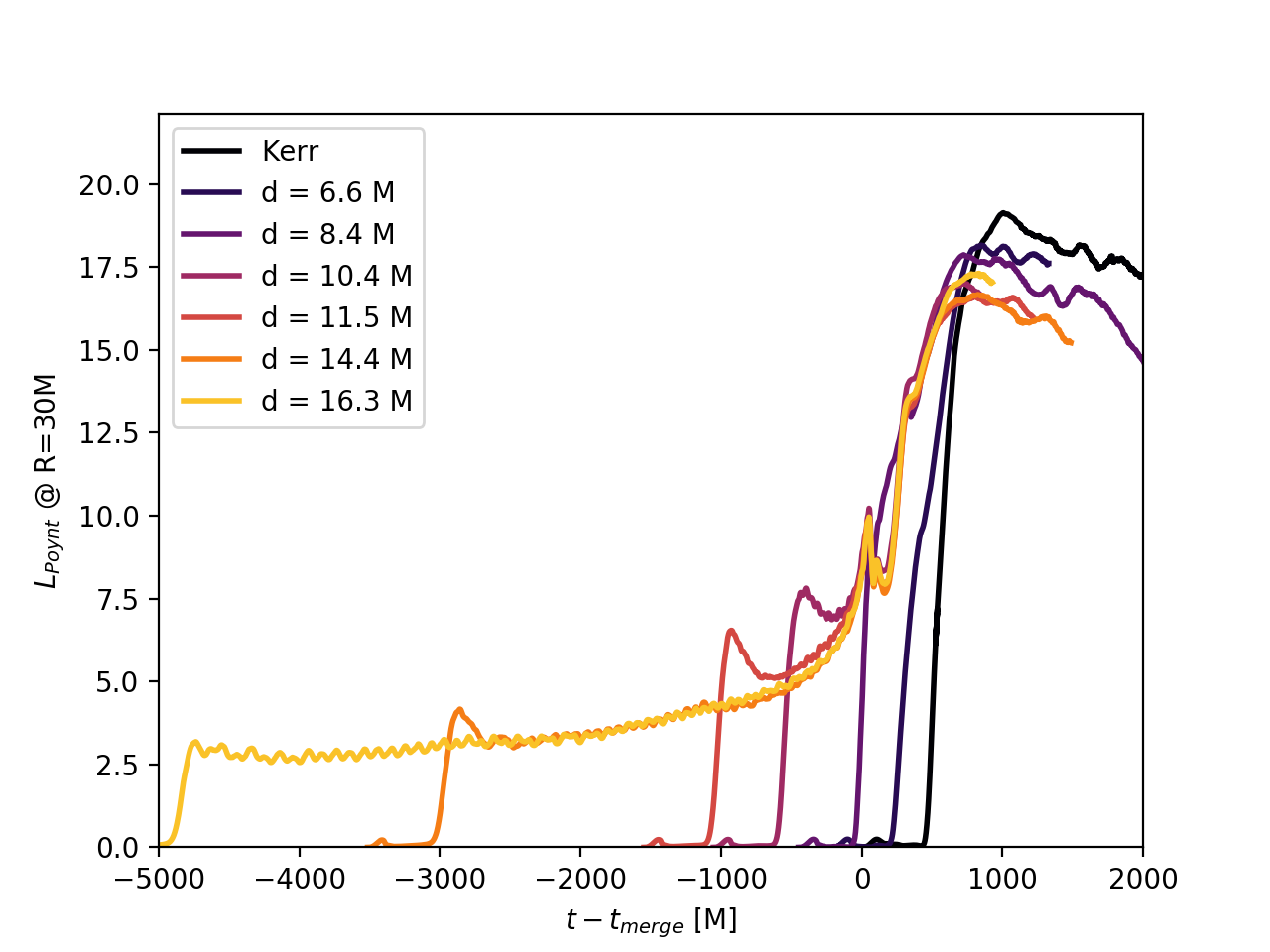}
  \caption{As in Fig.~\ref{fig:LEM_seps}, but with time axis aligned by merger time of the binary.
  ($1 \LEM = 5.867 \times 10^{44} \rhothir M_8^2 \, \UNITerg \, \UNITs^{-1}$.)
}
  \label{fig:LEM_seps_shifted}
\end{figure}

The blip (d) and the rise surrounding it do appear to be correlated
with the merger time. In
Fig.~\ref{fig:LEM_seps_shifted}, we realign the flux curves of
Fig.~\ref{fig:LEM_seps} by merger time $\tmerge$
(time when a common apparent horizon is first found;
 see Table~\ref{tab:merge_info}). It can be seen that the general trend
with larger separation has been to reveal a consistent pre-merger
portion of the flux. After an initial settling-in, the flux rises
slowly as the binary system inspirals. 

In Appendix \ref{apx:res_tests} we explore the robustness of this
result to changes in the extraction radius and to numerical resolution
changes. Overall the level and shape of the
curve in Fig.~\ref{fig:LEM_seps_shifted} provides a picture of the
time-dependence of the available jet power, which is robust at roughly
the ten-percent level.

\subsubsection{Magnetic Field Dependence of Poynting Luminosity}
\label{sssec:LEM_Bdep}

In the previous subsections we found a ``light curve'' for the time
dependence of outgoing Poynting flux for a canonical ambient fluid density and
aligned magnetic field strength of $b_0 = 0.1$.
However, it is natural to expect that features of EM flux will change as
the initial ambient field strength varies. In
previous studies carried out in the force-free limit \cite{Neilsen:2010ax,Palenzuela:2010xn}
the Poynting flux necessarily scaled with the square of the initial magnetic field strength. On the
other hand, if the matter flows play an important role in driving magnetic field development,
then we should expect a different scaling.

Here we investigate this issue by looking at several $d = 14.4M$ configurations that differ
only in their initial uniform magnetic field strength $b_0$.
The different field parameters are presented in Table~\ref{tab:Bdep_configs}, along with the resulting
Alfv\'{e}n speeds $\valf$.

\begin{table}\footnotesize
\centering
\caption{Initial uniform GRMHD field values for canonical $d = 14.4M$ configuration
         (\texttt{b1e-1}) (shown in bold face) and variants discussed in Secs.~\ref{sssec:LEM_Bdep}
         and \ref{sssec:LEM_scaling}.}
\begin{tabular}{l rrr|rrrrr}
\hline \hline
config               & $\rho_0$  & $b_0$     & $\kappa_0$ & $\press_0$      & $\epsilon_0$ &  $\umagofluid_0$ & $h^*_0$ & $\valf$ \\
\hline
\texttt{b1e-1}       & {\bf 1.0} & {\bf 0.1} & {\bf 0.2}  &  {\bf 0.2} & {\bf 0.6} & {\bf 5.0e-3}  & {\bf 1.81} & {\bf 0.074} \\
\hline
\texttt{b1e-2}       &  1.0      & 0.01      & 0.2	  &  0.2       & 0.6          & 5.0e-5  & 1.8 & 0.0075 \\
\texttt{b3e-2}       &  1.0      & 0.03      & 0.2	  &  0.2       & 0.6          & 4.5e-4  & 1.8 & 0.022 \\
\texttt{b3e-1}       &  1.0      & 0.3       & 0.2	  &  0.2       & 0.6          & 4.5e-2  & 1.89 & 0.22 \\
\texttt{b1e0}        &  1.0      & 1.0       & 0.2	  &  0.2       & 0.6          & 5.0e-1  & 2.8 & 0.60 \\
\hline
\texttt{b1e-1\_up}   &  100.0    & 1.0       & 0.0431     &  20.0      & 0.6          & 5.0e-3  & 1.81 & 0.074 \\
\texttt{b1e-1\_down} &  0.01     & 0.01      & 0.928      &  2.0e-3    & 0.6          & 5.0e-3  & 1.81 & 0.074 \\
\hline \hline
\end{tabular}
\label{tab:Bdep_configs}
\end{table}

\begin{figure}
  \includegraphics[trim=0mm 0mm 0mm 0mm,clip,scale=0.60]{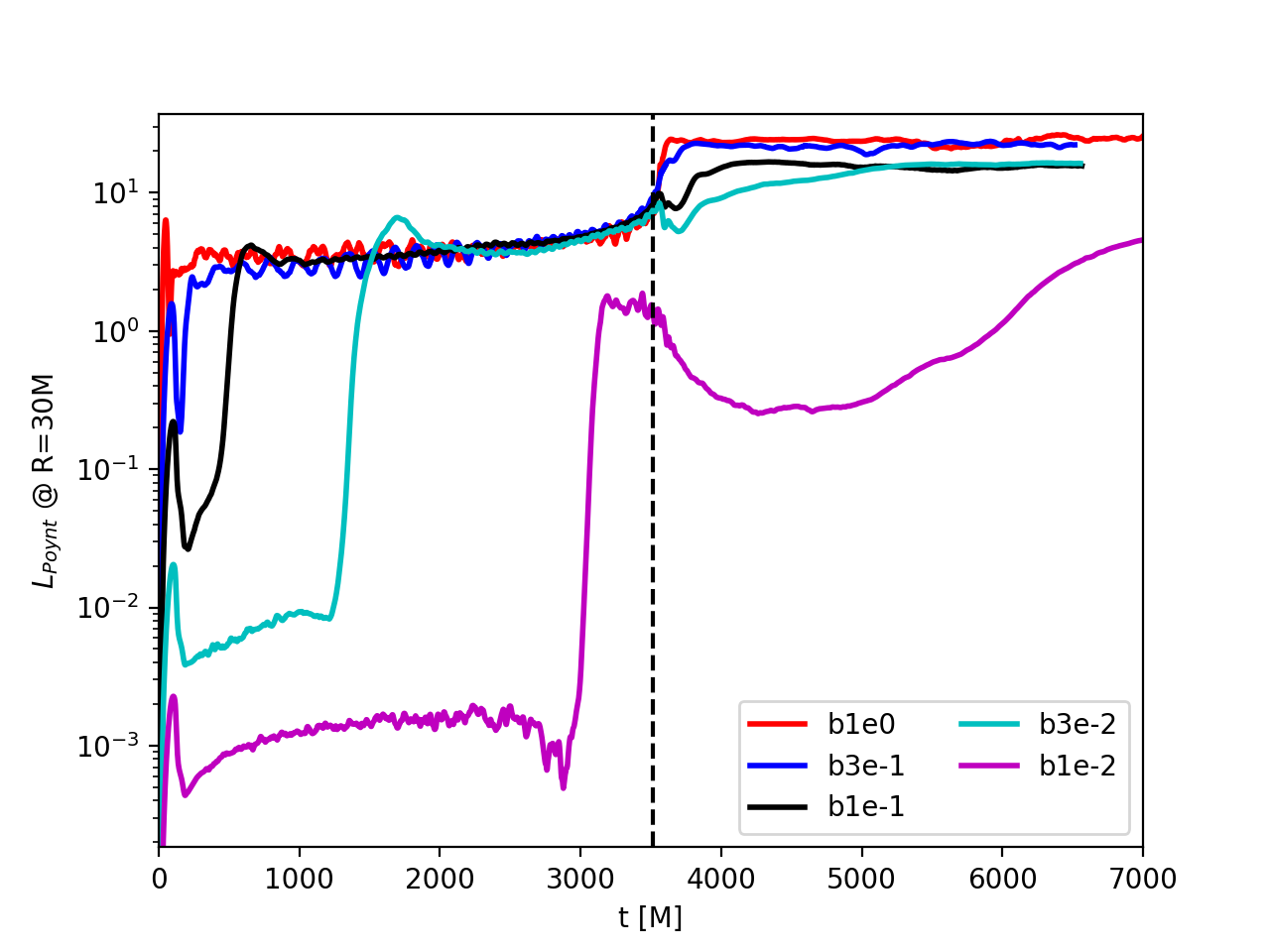}
  \caption{$\LEM$ for low-resolution $d = 14.4M$ configuration, for five different magnetic field
           strengths $b_0$, extracted at $R = 30 M$. The common merger time is indicated by the
           dashed vertical line.
  ($1 \LEM = 5.867 \times 10^{44} \rhothir M_8^2 \, \UNITerg \, \UNITs^{-1}$.)
          }
  \label{fig:LEM_R30_d14p4_allB}
\end{figure}

Figure~\ref{fig:LEM_R30_d14p4_allB} shows the resulting Poynting luminosities on a logarithmic scale.
While the flux in all cases exhibits a very small early amplification
(the ``initial-settling'' peak (a) in Fig.~\ref{fig:LEM_annotated})
whose timing is insensitive to field strength, the later rise to
levels observed during inspiral is significantly accelerated or
retarded relative to our canonical case, with stronger ambient fields
rising more quickly. The ``rise time'' is consistent with a feature
traveling outwards at the initial ambient Alfv\'{e}n speed $\valf$ (see Table~\ref{tab:Bdep_configs}), as
$\valf \propto b^2$ in non-magnetically dominated regions
(Eq.~\ref{eq:valf_def}).

More surprisingly, however, each configuration appears to reach the
\emph{same} level of Poynting luminosity during inspiral, regardless
of initial field strength (only the weakest of the five cases does not
share this common inspiral luminosity, presumably because $\valf$ is
too low for the disturbance to reach the observer at $R = 30 M$ before
merger).  This is important because insensitivity to details of
astrophysical conditions at the time of merger, as we seem to see
with magnetic field strength in this case, would be an important factor in
any potentially robust electromagnetic signatures of black hole
mergers.
  
\begin{figure}
  \includegraphics[trim=0mm 0mm 0mm 0mm,clip,scale=0.60]{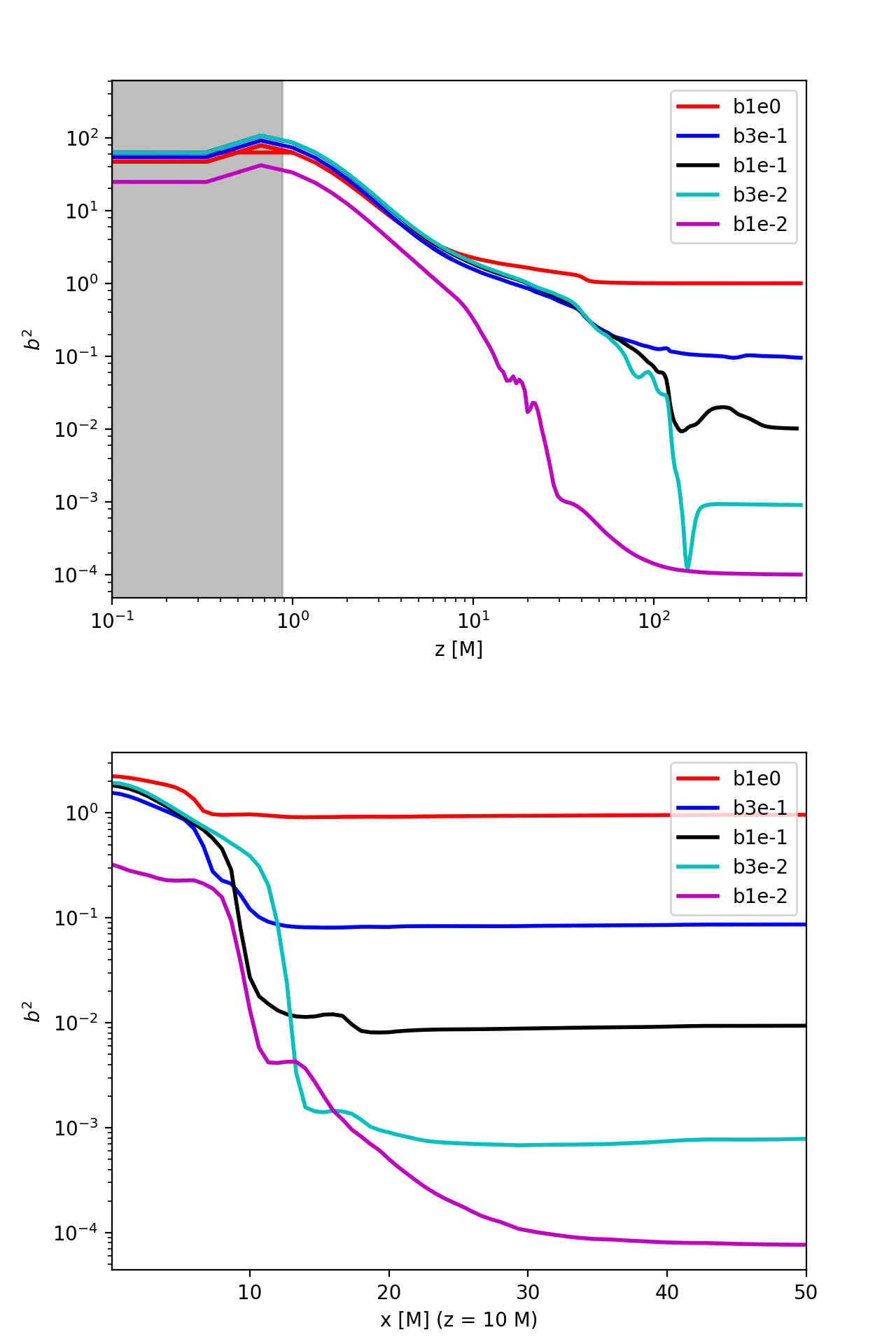}
  \caption{{\bf Top panel}: Evolved magnetic-field variable $b^2$ along the polar ($z$) axis at time $t = 5000 M$. The
           shaded region marks the interior of the black-hole horizon.
           {\bf Bottom panel}: $b^2$ along line parallel to $x$-axis, at $z = 10 M$.}
  \label{fig:smallb2_Z_LOWRES_allB_d14p4}
\end{figure}

To understand this apparent universality of the Poynting luminosity during inspiral, we next
analyze how the magnetic field is amplified in the vicinity of the binary. In the upper panel of
Fig.~\ref{fig:smallb2_Z_LOWRES_allB_d14p4}, we show the evolved
field $b^2$ as extracted along the orbital ($z$) axis for these configurations at time
$t = 5000 M$, about $1500M$ after merger. We see that, while $b^2$
asymptotes to its initial value far from the origin, the amplified
fields closer in tend to a common level. Indeed, within $\sim 10M$ of
the origin, the top four configurations are nearly indistinguishable,
reaching a common value of $b^2_{\rm max} \approx 100$ (similar to
what was reported in \cite{Giacomazzo:2012iv}).
The lower panel shows $b^2$ measured at the same time, but along a line parallel to the $x$
axis, at a height $z = 10 M$. As the configuration is highly axially symmetric around the orbital
($z$) axis by this time, this represents the general falloff of $b^2$ with distance from the
orbital axis. Grouping of the curves in the region $x<10M$ shows that
the consistency of $b^2$ along the axis is representative of the field
strength across most or all of the jet-like region.

This common magnetic field magnitude suggests a physical process in
which gravitationally driven matter flows drive up the magnetic field
to the point of saturation. The saturation likely reflects a point of
overall balance between magnetic pressure and gravitationally driven
matter pressure. Whatever the mechanism's details, its
effect is that the arbitrary initial fields are replaced by a
universal, magnetically dominated helical structure. The outgoing
Poynting flux thus also tends to a common level. We remind the reader
that our simulations scale with an arbitrary initial gas density
$\rho_0$. As the density increases, the magnetic field strength should
scale with $\rho^{1/2}$.

\subsubsection{Scaling Behavior of Luminosity}
\label{sssec:LEM_scaling}

In the matter-free simulation of black-hole mergers, the timescale and all observables (e.g.
gravitational-wave amplitude and frequency) scale with (or inversely to) the total mass $\MTOT$ of the
system; thus the same simulation can describe the merger of a stellar-mass system or a supermassive
one. 

\begin{figure}
  \includegraphics[trim=0mm 0mm 0mm 0mm,clip,width=0.5\textwidth]{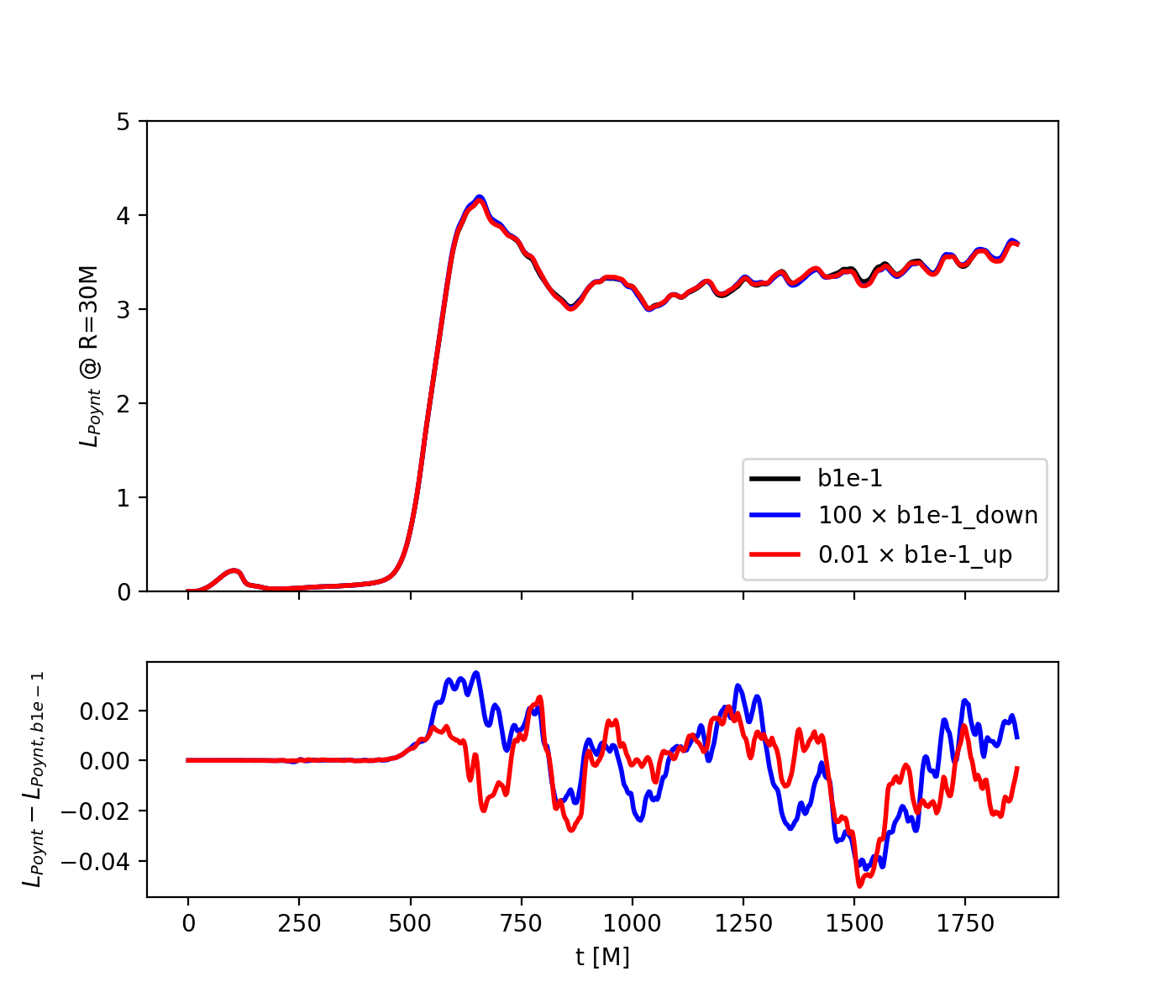}
  \caption{{\bf Top panel}: $\LEM$ for low-resolution $d = 14.4M$ configuration, for the three different choices of
           $\{ \rho_0, \press_0, b_0^2 \}$. {\bf Bottom panel}: difference between the canonical
           \texttt{b1e-1} configuration and each of the other two.}
  \label{fig:LEM_d14p4_scaling_alfven}
\end{figure}

The results of our GRMHD simulations in this work are not so trivially rescaled. In fact, for a
given binary mass $\MTOT$ (which sets the timescale), the Poynting luminosity scales cleanly only
with the combination $\{ \rho_0, \press_0, b_0^2 \}$. That is, if we wish to scale the magnetic field
strength $b^2_0$ by a factor $C$, then the same dynamics applies as long we also scale the initial
baryonic density $\rho_0$ and the pressure $\press_0$ by the same
factor.\footnote{Note that since the initial polytropic
  pressure-density relation \eqref{eq:polytrope_EOS} is nonlinear, the 
constant $\kappa$ must be adjusted to achieve the same scaling in $\press$ and $\rho$.}
The time-dependent Poynting luminosity is then $C$ times the original. We demonstrate in
Fig.~\ref{fig:LEM_d14p4_scaling_alfven} that this scaling is realized
computationally.

This scaling invariance should not be surprising, since the total stress-energy tensor
\eqref{eq:Tab_sum} is homogeneous in these three quantities. As long as gravitational effects from
the matter fields are not relevant then the dynamics will be independent of $C$. Consequently all
velocities, including for instance the Alfv\'en velocity \eqref{eq:valf_def}, are independent of
this collective rescaling.
If we further write the magnetic-fluid energy density ratio as $\umagofluid_0 \equiv b_0^2/(2 \rho_0)$, then
the uniform scaling performed in this section is equivalent to scaling the initial fluid density
$\rho_0$ while keeping the specific internal energy $\epsilon_0$ and the energy-density ratio
$\umagofluid_0$ constant.

For a fixed fluid density $\rho$, the luminosity scales with volume divided by time. In geometric
units, this ratio scales as $\MTOT^2$.
Thus the luminosity satisfies the scaling relation
\beq
\LEM(t) = \rho_0 \MTOT^2 F(t/\MTOT;\epsilon_0,\umagofluid_0) \label{eq:LEM_rhodep},
\eeq
where $F(t/\MTOT;\epsilon_0,\umagofluid_0)$ is a dimensionless function of time.
In the context of EM counterparts this scaling differs from many other emission models that scale
roughly with $\MTOT$, as in Eddington-limited accretion. Note that our study does not model EM
radiation feedback, which would control an Eddington-limited process \cite{Krolik:2009hx}.

The choice of initial density, however, can itself be influenced by the total mass of the system.
For instance, consider the geometrically thick accretion disks
investigated by \cite{Gold:2013zma}, $\rho_0 \sim \MTOT^{-1}$.
In such a system, the Poynting luminosity \eqref{eq:LEM_rhodep} will
scale linearly with $\MTOT$. Our results above indicate furthermore
that $F(t/\MTOT;\epsilon_0,\umagofluid_0)$ is effectively independent
of $\umagofluid_0$ over a significant range of magnetic field
strength. Of course, in the limit of extreme magnetic dominance, we
expect the FFE description to apply, where density can be assumed to
be irrelevant, and the luminosity scales with magnetic field squared.

At least for the simple class of astrophysical scenarios covered in
our simulations we conclude that Poynting flux---as a time-dependent
driver for jet energy---is largely independent of several
astrophysical details, particularly magnetic field strength, up to a
simple scaling. Next we consider the relation of its time dependent
behavior to orbital dynamics.

\subsubsection{Relation between Poynting luminosity and Orbital Motion}
\label{sssec:LEM_vdep}

Several numerical~\cite{Palenzuela:2010xn,Palenzuela:2010nf,Neilsen:2010ax,Paschalidis:2013jsa} and
analytical~\cite{Lyutikov:2011vca,McWilliams:2011zi,Penna:2015qta,Morozova:2013ina} studies have
investigated how even non-spinning black holes in an orbital configuration can generate Poynting
luminosities in the limit of force-free MHD through a process similar to the Blandford-Znajek
mechanism~\cite{Blandford:1977ds} for jets powered by a black hole. In Blandford-Znajek, the
twisting of magnetic field lines in interaction with a spinning black hole converts kinetic energy
to jet power. Before merger, however, the dragging of black holes through the ambient field
similarly converts kinetic energy to jet power. While there are differences in the computed
efficiency of this conversion, a general picture emerges that (for
nonspinning black holes in the inspiral phase) the Poynting luminosity scales as
\beq
  L_{\rm FFE, insp}\sim v^2 B^2 M_{\rm BH}^2.
  \label{eq:LEM_vdep_FFE}
\eeq
A difference with our simulations is that our black holes do not orbit
in a magnetically dominated, force-free environment. Here we investigate whether a similar
velocity scaling still holds, analyzing data in the case with the longest inspiral: $d= 16.3M$.

We derive instantaneous BH velocity data from the motion of the BH
horizons given by our apparent horizon finder. While these velocities
are not gauge-invariant, in practice they are reliable after an
initial settling-in time of $\sim 50M$ and before the formation of the
common horizon at merger.

\begin{figure}
  \includegraphics[trim=0mm 0mm 0mm 0mm,clip,scale=0.60]{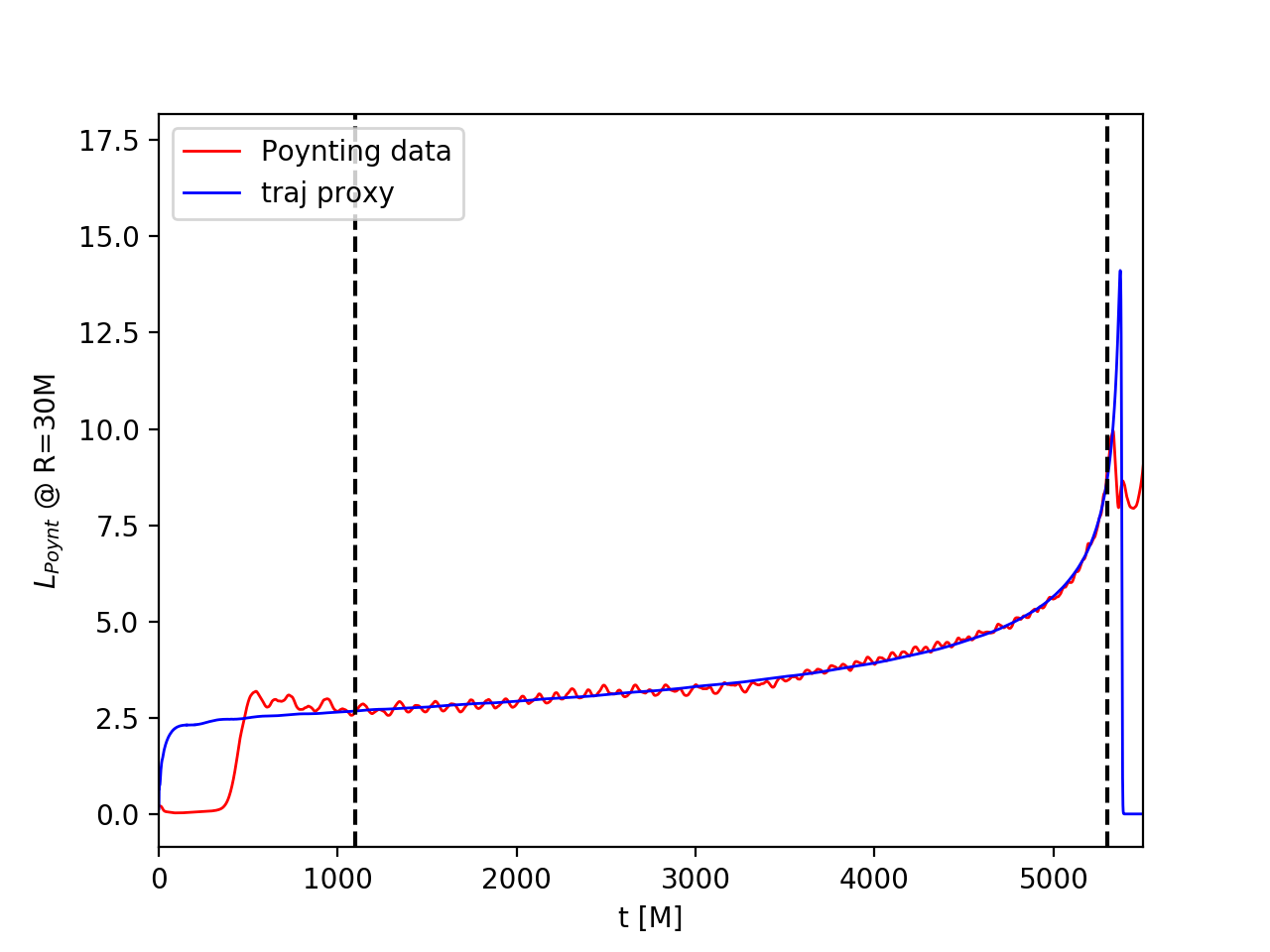}
  \caption{Best scaling of form $A v^n$ to match puncture speed $v$ with $\LEM$ as extracted at $R = 30M$.
  The dashed vertical lines indicate the beginning and end of the fit region.}
  \label{fig:LEM_d16p3_trajectory_comparison}
\end{figure}

Complicating this issue is the time lag between the source motion and the resulting Poynting flux
present in fields measured farther out. In Fig.~\ref{fig:LEM_d16p3_trajectory_comparison}, we show the
best fit between $\LEM$ as measured at $R = 30 M$ and the measured speed, assuming
$\LEM(t) = A v^n$, where $v$ is measured at time $t$ offset by some fixed time $\Delta$,
representing propagation from the strong-field region of the BHs to the extraction radius $R$. The
best-fit parameter values are $\Delta = 100 M$, $A = 870$, and $n = 2.7$, based on $\LEM$ over
an inspiral ``segment'' beginning once $\LEM$ has settled down into the inspiral regime, and ending
at the merger blip (times indicated by vertical dashed lines in the Figure).

The best-fit value $\Delta = 100 M$ is consistent with $\LEM$ propagating from the strong-field
region out to $R = 30 M$ at an effective speed of $v_{\rm prop} \approx 0.33 c$.
In principle, if we know that the Poynting flux is always propagating outward at a well-defined
Alfv\'{e}n speed $\valf$, we can derive the necessary time shift $\Delta$ from that. However,
$\valf$ changes with time and position --- increasing as the underlying $b^2$ grows and $\rho$
declines --- and such a detailed analysis is beyond the scope of this paper.

Thus we can deduce that for a fixed initial field configuration, during the inspiral phase the
Poynting luminosity depends on the orbital motion as
\beq
\LEMinsp(t) \approx A \rho_0 \MTOT^2 v^n(t_{\rm ret}), \label{eq:LEM_vdep}
\eeq
with best-fit values $A=870$, $n=2.7$ and $t_{\rm ret}=t-100M$.
While this result is derived from just one of our runs, we have established above that the inspiral
portion of our runs yields similar results independent of the magnetic field strength and of the
initial orbital separation at which we set the plasma to be at rest in our numerical coordinates.
In our case, it is the \emph{fluid density} that scales
the luminosity, and that seems to regulate the magnetic field strength.
Independent of the observed invariance to initial magnetic field
strength, comparison with \eqref{eq:LEM_vdep_FFE} reveals an
enhanced brightening as the velocity increases.
One interpretation of this
enhancement would be that a mechanism similar to that observed in the
FFE studies is also generating power in our studies. However, in our
cases the magnetic field strength grows on approach to merger,
due to the accretion of gas and thus piling up of field lines near the horizon.

\subsubsection{Formula for Luminosity in Magnetized Plasma}
\label{sssec:LEM_formula}

Armed with the observations of the previous subsections, we can
summarize our results for Poynting luminosity of the binary at a
representative reference point in its ``inspiral'' phase, and at
peak. Given that the BH orbital speed increases only gradually even
late in the inspiral, we choose a representative speed $v_{\rm insp} =
0.13 c$ (this corresponds to a puncture separation of $d \approx
12.2M$, about $2000M$ before merger.). Then from \eqref{eq:LEM_vdep}, we obtain for the
inspiral
\begin{align}
\LEMinsp & \approx 3.55 \, \rho_0 \, \MTOT^2 \nonumber \\
         & = 3.55 \, \eta_{\rm cgs} \, \rhothir \, M_8^2 \, \UNITerg \, \UNITs^{-1} \nonumber \\
         & \approx 2.1 \times 10^{45} \, \rhothir \, M_8^2 \, \UNITerg \, \UNITs^{-1}, \label{eq:inspiral_formula_cgs}
\end{align}
where we use the conversion factor $\eta_{\rm cgs}$ from Eq.~\eqref{eq:code_to_cgs} to convert from
code units to cgs.

Judging from Fig.~\ref{fig:LEM_seps_shifted}, the post-merger peak of $\LEM$ is around $17$ in code
units for our canonical case. However, this is derived from a set of simulations carried out at
modest resolution ($M/48$). As noted in Appendix~\ref{apx:res_tests}, post-merger values
of $\LEM$ increase somewhat with resolution. If we round up so that the peak Poynting
luminosity is $\LEMpeak \approx 20$ in code units, we find
\begin{align}
\LEMpeak &\approx 20 \, \rho_0 \, \MTOT^2 \nonumber \\
         &\approx 1.2 \times 10^{46} \, \rhothir \, M_8^2 \, \UNITerg \, \UNITs^{-1}. \label{eq:peak_formula_cgs}
\end{align}
This can be combined with the mass accretion rate found in
Sec.~\ref{ssec:results_mdot} to estimate a Poynting radiative efficiency around the merger:
\beq
\epsilon_{\rm EM} \equiv \frac{\LEMpeak}{\Mdot c^2} \approx 0.22. \label{eq:poynting_efficiency_MHD}
\eeq

\subsubsection{Comparison with Previous Results}
\label{sssec:LEM_FFE}

In the previous subsection we quantified potential Poynting-flux-powered emissions, synthesizing
the results obtained from our GRMHD simulations of mergers with initially non-magnetically-dominated plasmas.
We can compare these with the results of previous GRFFE studies~\cite{Neilsen:2010ax,Palenzuela:2010nf,Palenzuela:2010xn}
and with previous GRMHD studies of mergers in circumbinary disk configurations~\cite{Gold:2013zma,Gold:2014dta,Farris:2012ux}.

Quantitative comparisons depend on assumptions
about the astrophysical environment. Leaving aside details of the matter distribution, the
environment of our simulations is characterized by a scalable initial gas density relative to a
reference density of $\rhothir=1=\rho_0/(10^{-13} \, \UNITg \, \UNITcm^{-3})$ with an initially uniform
poloidal magnetic field. In these units the magnetic field strength of our canonical
configuration was $B_4=B/(10^4 \, \UNITG)=0.34 \rhothir^{1/2}$, but we found that the Poynting flux
is minimally changed if the magnetic field strength is varied by an order of magnitude either up or
down. Thus for fixed black hole mass, our overall result for $\LEM$ simply depends linearly on
initial density.

Our study resembles previous GRFFE simulations in that both assume an
initially uniform large-scale poloidal magnetic field. As we have
noted, the magnetic field structures and the velocity dependence
on approach to merger strongly resemble GRFFE results. However, GRFFE
results apply in the regime where the fluid is magnetically dominated
and are thus independent of density. Instead, the relevant scale
parameter for the environment is the magnetic-field energy
density. Those authors suppose an astrophysically motivated reference
scaling of $B=10^4G$.  Despite the scaling differences, we can
nonetheless compare with our results at particular magnetic-field and
fluid density values.

Since the previous GRFFE simulations involved only relatively brief
simulations, it makes more sense to compare peak levels of
Poynting luminosity. In the figures and discussion of
Refs.~\cite{Neilsen:2010ax,Palenzuela:2010nf,Palenzuela:2010xn} the
Poynting luminosity tends to rise to a brief peak and
then to quickly fall off to a level appropriate for the final spinning
black hole, while our luminosities stabilize closer to their peak
levels at late times. Taking this and differences in the various FFE
papers into account we estimate a peak level from these publications,
which can be compared to our Eq.~\eqref{eq:peak_formula_cgs}, of
\beq
  L_{\rm FFE, peak}  \approx 3 \times 10^{43} {B_4}^2 \, {M_8}^2 \, \UNITerg \, \UNITs^{-1}, \label{eq:FFE_peak_formula_cgs}
\eeq
reliable within a factor of two.
At nominal values the previous GRFFE studies yield a peak
Poynting luminosity level about $400$ times smaller than our nominal
result, but the assumptions about the astrophysical environments are
not quite consistent; in our simulations, the environment is not
initially magnetically dominated.

Using the above estimates and expressions, can we then find the value
of $B_4$ for the GRFFE environment in Eq.~\eqref{eq:FFE_peak_formula_cgs} to achieve
the same Poynting luminosity that we see in our canonical case? The
answer is $B_{*4}\approx 20$ plus or minus $50\%$. Converting
to the units of our simulations using Eq.~\ref{eq:convert_B} for the
relevant case $\rhothir=1$, this corresponds to
$\umagofluid=b_{*}^2/(2\rho_0)\approx18$, which, appropriately enough,
is higher than the initial magnetic field strengths of any of our
simulations.

We note that the equivalent value $b_{*}^2 \approx 35$ is close to the \emph{evolved} $b^2$ values
seen near the post-merger black hole in our simulations (which we found to be roughly independent
of initial field strength; see Fig.~\ref{fig:smallb2_Z_LOWRES_allB_d14p4} and discussion in
Sec.~\ref{sssec:LEM_Bdep}).

This suggests the following shorthand description of the comparison between the
results of our simulations and previous GRFFE results: The expression
\eqref{eq:FFE_peak_formula_cgs} for the GRFFE Poynting luminosity
gives an approximately correct description of the our initially
matter-dominated GRMHD simulations if, in place of the initial
magnetic field strength $B_4$, the dynamically driven
magnetic field strength found near where the jet meets the horizon is
used instead.

We can also compare with Poynting luminosities from previous binary
black hole simulations with matter initially structured in a
circumbinary disk. Using the code on which \IGM{} is based,
Refs.~\cite{Gold:2013zma,Gold:2014dta} bring a $\Gamma=4/3$, non-self-gravitating circumbinary disk
with a poloidal magnetic field to quasi-equilibrium by allowing an equal-mass BBH to orbit at fixed
separation
for $\sim$ 45 orbital periods. To ensure quasi-equilibrium could be
established with reasonable computational cost, the disk was assumed
to be thick ($H/R\sim 0.3$) so that the MHD turbulence
(magneto-rotational instability) driving the accretion could be
adequately resolved.
Beginning from a point about $700M$ before merger, the
binary was then allowed to inspiral and merge, solving the full set of
general relativistic field equations for the gravitational fields and
the equations of GRMHD for the (non-self-gravitating) disk
dynamics. 

A quantitative comparison of our results with results of
Refs.~\cite{Gold:2013zma,Gold:2014dta} for circumbinary disks is
challenging. First, we can only compare with their fixed choice of
magnetic field configuration. Given that we observe some degree of
insensitivity to the initial magnetic fields chosen, we will suppose
that their field is within a broadly comparable range, noting that
their simulations also include regions of gas and magnetic pressure
dominance. More fundamental are the density scales near the horizons
that power Poynting luminosity. While such densities in our simulations
span roughly an order of magnitude,
densities in the circumbinary disk simulations span many more. Thus
there is no clear way to define a common density as a point of
reference for the two studies. Instead, we make a comparison of
Poynting luminosities normalized by the mass accretion rate (i.e.,
``Poynting luminosity efficiency'') during and after merger as an
indicator of the supply of gas in the vicinity of the black holes.

The mass accretion rate in Ref.~\cite{Gold:2014dta} varies significantly before merger, but settles
to a value near $0.1$ in their units (see their Fig.~3).
Scaled by this value, their Poynting luminosity efficiency is close to $\epsilon_{\rm EM} \equiv \LEM/\dot M\approx0.01$ near merger,
growing by about a factor of 5 during the subsequent period of $1000M$.
Their peak efficiency is reached at a similar time after merger as in our simulations,
but remains smaller than our peak value
(Eq.~\eqref{eq:poynting_efficiency_MHD}) by a factor of a few.

\subsection{Simulating Direct Emission from Merger}
\label{ssec:results_pandurata}

To this point, we have focused primarily on the Poynting flux as a
proxy for EM power from the merging black holes. However, Poynting
flux alone is not directly observable; we interpret it as a power
source for EM emissions downstream along the jet. An alternative
mechanism for EM emissions is direct emission from the plasma fluid.

In our simulations the lack of a
realistic equation of state or of any radiative cooling
mechanism for the gas makes it difficult to produce a reliable
prediction for the actual EM emission. Further,
our initial conditions of uniform density and magnetic
fields do not capture astrophysical details of the full
system that may also contribute to EM emission.

We have carried out a simplified
calculation of the EM luminosity generated during the inspiral and
merger simulation. To do so, we have used a new version of the Monte
Carlo radiation transport code \Pan{} \cite{Schnittman:2013lka},
revised to allow for arbitrary spacetime metrics. While the \IGM{}
simulations generate a real dynamic spacetime by solving
Einstein's equations numerically, for this toy emission model we employ
a simplified version of the metric that can be calculated efficiently
by \Pan{} as a post-processor of the MHD data. As
described in \cite{Schnittman:2017nhg},
the binary four-metric can be instantaneously described by a three-metric $\gamma_{ij}$, lapse
$\alpha$, and shift $\beta^i$, according to:
\beq
g_{\mu \nu} = \begin{pmatrix}
-\alpha^2 + \beta^2 & \beta_j \\
\beta_i & \gamma_{ij}\\
\end{pmatrix}\, .
\eeq
Following \cite{Campanelli:2006uy}, we use $\alpha = 2/(1+\psi^4)$,
$\beta_j =0$, and $\gamma_{ij}= \delta_{ij}\psi^4$. The conformal
factor $\psi$ is given by
\beq
\psi = 1+ \frac{m_1}{2r_1}+\frac{m_2}{2r_2}\, ,
\eeq
with $r_1$ and $r_2$ being the simple Cartesian distances between the
spatial coordinate and the primary/secondary masses. For the
Christoffel-symbol components $\Gamma^{\rho}_{\mu \nu}$ we take 
the spatial and temporal metric derivatives analytically based
on the puncture trajectories calculated by the apparent horizon
finder used in our GRMHD simulations. One advantage of using this
simplified metric is that we can easily calculate the photon
trajectories ``on the fly'' and thus do not need to rely on the fast
light approximation used by many ray-tracing codes. 

Even though \Pan{} uses a slightly different metric than that of the GRMHD
simulations, the qualitative properties of the spacetime are expected
to be very similar. We can avoid some potential numerical problems
by normalizing the \IGM{} fluid 4-velocity
everywhere by using the coordinate 3-velocity from \IGM{} and then
using the analytic metric to solve for $u^t$ via $g_{\mu \nu}u^\mu u^\nu=-1$. 

Given the fluid velocity at each point and for each data snapshot, a
local tetrad can be constructed as in \cite{Schnittman:2013lka}, from which
photon packets are launched and then propagated forward in time until
they reach a distant observer or are
captured by one of the black holes. Those that reach the observer are
combined to make images, light curves, and potentially spectra.
We ignore
scattering or absorption in the gas, so that all photon packets
travel along geodesic paths.

\begin{figure}[h]
\begin{flushright}
\includegraphics[width=0.5\textwidth]{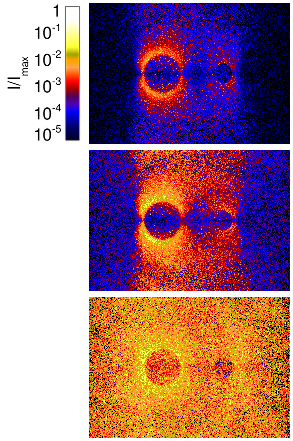}
\end{flushright}
\caption{Snapshots from \Pan{} post-processing of the simulation data
  at a separation of $10M$ (about $1000M$ before merger), viewed by an observer edge-on to the
  orbital plane. {\bf Top panel}: thermal synchrotron emission; {\bf middle panel}:
  magnetic contribution only ($\propto B^2$); {\bf bottom panel}: gas contribution
  only ($\propto \rho T$).
}
\label{fig:Pan_images}
\end{figure}

One of the challenges with this approach is the inherent uncertainty of
what emission mechanism is most appropriate, and even then, the
electron temperature $T_e$ is not known explicitly from the simulations, so it
can only be approximated with an educated guess. For this paper, we focused on a
single simplified emission model of thermal synchrotron, where the
emissivity is isotropic in the local fluid frame with bolometric power
density given by
\beq
P_{\rm syn} = \frac{4}{9}n r_0^2c\beta^2\gamma^2B^2 , \label{eqn:P_synch}
\eeq
with $r_0$ the classical electron radius, $n$ the electron number density,
$\beta \equiv v/c$, and $\beta^2\gamma^2 \approx T_e/m_e$
(see, e.g. Chap. 6 of \cite{RybickiLightman_1979}).
We use
the magnetic field strength and fluid density specified by \IGM{}, along
with the code-to-cgs conversion described above. We estimate the
electron temperature from the simulation pressure, assuming a
radiation-dominated fluid with $\press = aT_e^4$, reasonable for the
$\Gamma=4/3$ polytrope used here. Thus the synchrotron power scales as
\beq
P_{\rm syn} \propto B^2 \rho^{4/3} \propto \rho_0^{7/3},
\eeq
since $B^2 \sim \rho$.

In the top panel of Fig.~\ref{fig:Pan_images} we show the observed
synchrotron intensity on a log scale for a single snapshot of \IGM{}
data when the binary separation is $10M$. The observer is located
edge-on to the orbital plane and the black hole on the left is moving
towards the observer, resulting in a special relativistic boost.

In an attempt to understand the features seen in Fig.~\ref{fig:Pan_images}, we repeat the \Pan{} calculations with two other
emissivity models, in one case focusing just on the contribution from
the magnetic field, and in the other case on the electron density and
temperature. As can be seen in Fig.~\ref{fig:rho_d14p4_t2400}, the
gas forms two very small, thin disks with magnetically dominated
cavities above and below each black hole. From this picture alone, it
is not clear where most of the synchrotron flux might
originate. 

However, when comparing the three panels of Fig.~\ref{fig:Pan_images}, we see that the gas
contribution is almost uniformly distributed, and even the thin disks evident in
Fig.~\ref{fig:rho_d14p4_t2400} are almost indiscernible when all the
relativistic ray-tracing is included. The reason for this is
two-fold. First, the disks are quite small in extent, and the gas is
moving almost entirely radially, so the emitted flux is beamed into
the horizon, and thus the disks themselves are not clearly
visible in the ray-traced image. Second, the overdensity of gas in the
disks is only a factor
of a few or at most ten greater than the background density. On the
other hand, in the funnel regions, $B^2$ can be more
than four orders of magnitude greater than the ambient or initial
pressure, yielding much more significant spatial variations. Thus the
synchrotron image (top panel) most closely traces the magnetic field,
with a slight enhancement of emission where the gas density and
temperature rise near the black holes. 
  
\begin{figure}[h]
\includegraphics[width=0.5\textwidth]{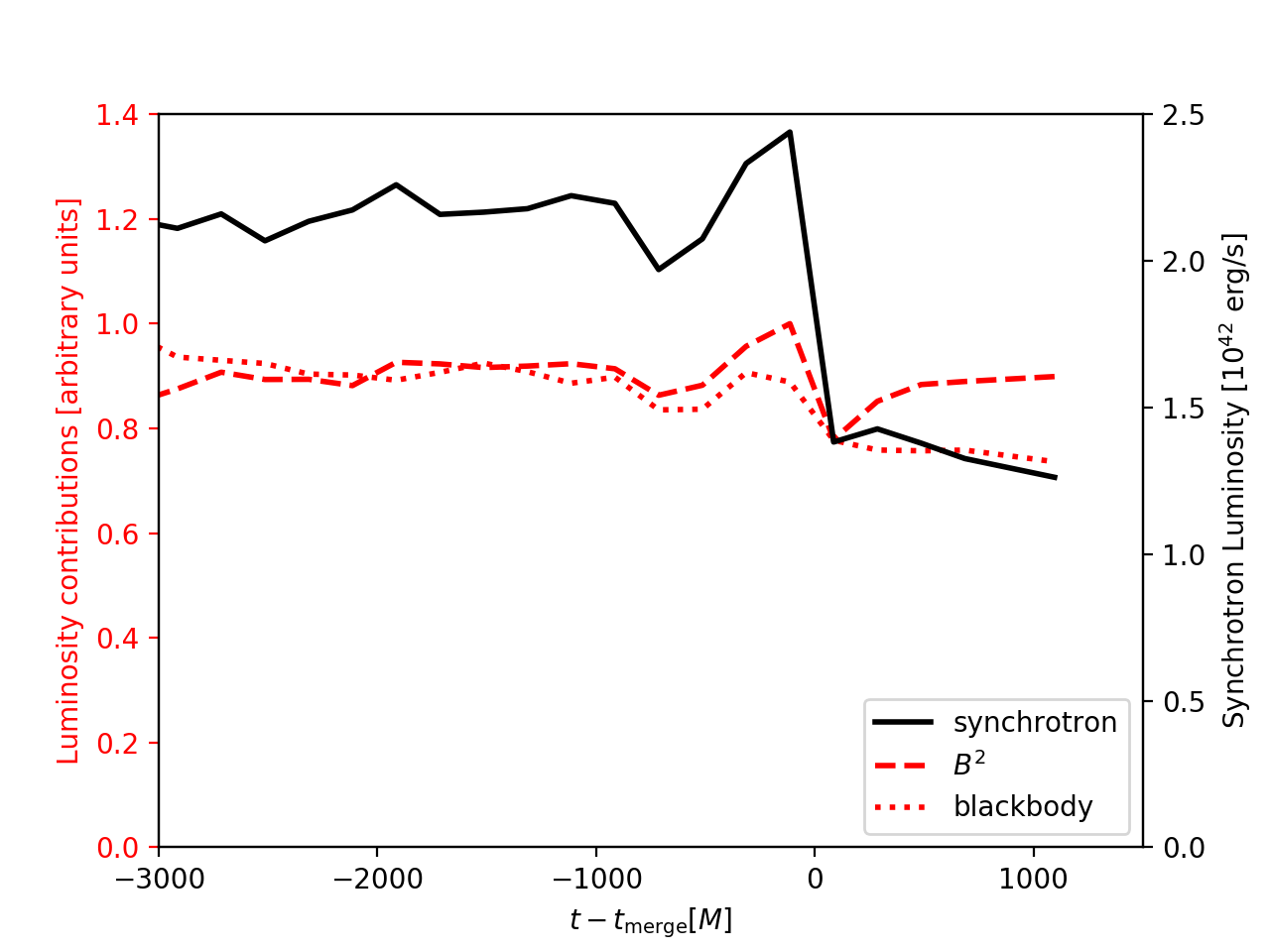}
\caption{Bolometric luminosity generated in the region $r<30M$ for the \texttt{X1\_d14.4}
         configuration, assuming the canonical initial density
         $\rho_0 = 10^{-13} \, \UNITg \, \UNITcm^{-3}$. We model local synchrotron emissivity, also showing
         the development of two contributing components as described in the text.}
\label{fig:lightcurves}
\end{figure}

In Fig.~\ref{fig:lightcurves} we show the light curve generated by
synchrotron emission along with analogous traces computed from the density and magnetic-field
components for the \texttt{X1\_d14.4} configuration.
To calculate these curves, millions of photons must be launched
at each time step, so for efficiency's sake, we use a relatively
coarse time sampling of $200M$. We only consider emission from inside
$r<30M$, consistent with the Poynting flux extraction radius.

Figure~\ref{fig:lightcurves} shows that, unlike the Poynting flux, the
locally generated EM power is nearly constant throughout the inspiral
leading up to merger. There is a small burst of luminosity preceding merger,
followed by a dip of almost $50\%$ for the synchrotron light curve,
but the other models show almost no discernible sign of the merger at
all. The dip is caused by the sudden expansion of the horizon volume
at merger, rapidly capturing the gas with the highest temperature and
magnetic field. 

Another curious result of the \Pan{} calculation is that,
for a single snapshot, there is very little difference in the flux
seen by observers at different inclination angles or azimuth (of order
$\sim 10\%$), suggesting that variability in the EM light curves on
the orbital time scale will be minimal. 

In principle, \Pan{} can also be applied to study the spectra of EM emissions including effects,
such as inverse-Compton scattering as photons interact with hot atmospheric plasma, that have been
found to be important in modeling black hole accretion disk spectra \cite{Schnittman:2012yn}. Our
present simulations, however, do not provide a realistic treatment of atmospheric densities and
temperatures. Future studies with more detailed physics may reveal more interesting time
development in spectral features of the emission.

The above simplifications and caveats mean that we cannot make robust statements about the observability of direct emission.
However, based on our optically thin synchrotron emission model, the direct emission luminosity
is orders of magnitude lower than that of the Poynting flux.
In addition, the synchrotron flux is roughly isotropic, while
significant beaming is observed in Poynting flux. There is no
contradiction in these measures; Poynting luminosity may manifest as
photons far downstream from the GRMHD flows, whereas these direct
emission estimates
originate in regions of high fluid density and magnetic field strength in strong-gravitational-field zones.

When comparing these direct emissions with results from circumbinary disk simulations,
the most similar simulation is in \cite{Gold:2013zma,Gold:2014dta}. They estimated a form of
direct emission, derived from a cooling function based on hydrodynamic shock heating. The
implied cooling luminosity was more than an order of magnitude \emph{larger} than the Poynting
luminosity, while our results suggest that Poynting luminosity is larger than direct synchrotron
emission, at least for the canonical density of $10^{-13} \, \UNITg \, \UNITcm^{-3}$. We have not
incorporated a similar cooling function for a more direct comparison, though we note that our gas
does not exhibit strong shocks.

\section{Conclusions and Future Work}
\label{sec:conclusions}

To deepen our understanding of the interplay of gravity, matter, and
electromagnetic forces in the vicinity of a merging comparable-mass
black-hole binary, we have carried out a suite 
of equal-mass non-spinning BBH merger simulations in uniform plasma
environments.  We considered two classes of potential drivers for
electromagnetic emissions, primarily focusing on the development of
Poynting flux, which may drive a jet, but also considering direct emissions
from the fluid.

We conducted simulations covering a range of nearly uniform density, low-velocity distributions of
hot gas with a significant but not dominant poloidal magnetic field. Based on these we find that
the Poynting luminosity grows on approach to merger (roughly with a power of orbital velocity
$v^{2.7}$), leveling off at a steady value after merger. The level and time development of the
Poynting luminosity is largely independent of the initial magnetic field strength and
not strongly dependent on initial pressure or small changes in fluid
configuration, scaling overall with density and the square of black
hole mass. Consistent with this we find that the central magnetic
field strength is largely independent of the initial field strength,
regulated by the gas flow. 
We further find that the coalescence yields a Poynting efficiency of 0.04 -- 0.22 between late
inspiral and merger.

These findings, using the new \IGM{} code, both confirm and extend our
earlier GRMHD results obtained with the \Whisky{} code
\cite{Giacomazzo:2012iv}, and form a bridge to complementary results
from GRFFE codes, in which the plasma is assumed completely
magnetically dominated.

Overall consideration of our results with those of previous GRFFE studies suggests a consistent
picture where below a transition point near ${B_4}^2\sim400\rhothir$, the gas flow dominates and
peak Poynting flux is described by our expression \eqref{eq:peak_formula_cgs}. Beyond this point
the plasma is magnetically dominated and the GRFFE results, summarized in
\eqref{eq:FFE_peak_formula_cgs} should apply.

To complement Poynting luminosity investigations, we also consider direct synchrotron emission
from the plasma in the strong-gravity region near the black holes.
To explore the
time-dependent bolometric luminosity in this scenario, we employ a new
version of the \Pan{} code to propagate photons through
the \IGM-generated MHD fluids (in post-processing) and generate
time-dependent EM flux.
Contrary to the Poynting flux analysis we do not find growth in the
synchrotron emission on approach to merger. Instead, the luminosity
remains steady until it drops to a slightly lower level after merger.
Note however that the physical processes behind the two emission mechanisms are mostly independent.
Poynting luminosity is due to the highly twisted and amplified magnetic fields in the larger funnel
region around the orbital axis, and is expected to accelerate charged particles to produce jet-like
behavior leading to EM emission farther downstream; while the direct
emission considered here is due to the plasma itself in the more
immediate vicinity of the black holes, both before and after merger.

These results provide clues about the physical processes which may
drive electromagnetic counterparts to massive black hole
mergers, which future GW instruments such as LISA may observe. However
limitations to these studies prevent more
definitive counterpart predictions. As with many similar studies, our study
assumes a large-scale orbit-aligned magnetic field that might
approximate the local astrophysical environment near a massive BBH. While some
of our results are independent of the level of this field, it provides
an asymptotic field structure that we have not strongly
justified.

Similarly, while we find that EM emissions are sensitive to ambient gas
density, it is unclear how well our very simplified gas distribution,
lacking angular momentum support, stands in for real flows from a
larger available gas reservoir, such as a circumbinary disk. Our
simulations also lack dynamical effects from radiation flows including
radiative cooling effects. These limitations will motivate our future
work.
With more realistic gas distributions in place, we will also investigate the effects of less
symmetrical BH systems, including merger recoils~\cite{Zanotti:2010xs}.

\acknowledgments

BJK, ZBE, JGB, and JDS acknowledge support from the NASA grant ATP13-0077.
The new numerical simulations presented in this paper were performed
on the Pleiades cluster at the Ames Research Center, with
support provided by the NASA High-End Computing (HEC) Program, as well
as on West Virginia University’s Spruce Knob supercomputer, funded by
NSF EPSCoR Research Infrastructure Improvement Cooperative 
Agreement \#1003907, the state of West Virginia (WVEPSCoR via the
Higher Education Policy Commission), and West Virginia University.

\appendix

\section{Relation of $\Szdom$ to Electromagnetic Flux}
\label{apx:LEM_Sz10}

The quantity $\Szdom$ used in the main text is closely related to the EM luminosity calculated by \cite{Neilsen:2010ax,Palenzuela:2010xn}
in terms of the ``outgoing'' Newman-Penrose \cite{Newman:1961qr} EM radiation scalar $\Phi_2 = F_{ab} n^a \bar{m}^b$
\beq
\LEM = \frac{d E_{\rm EM}}{dt} = \lim_{R \rightarrow \infty} \oint \frac{R^2}{2 \pi} |\Phi_2|^2 d\Omega.
\eeq
The modulus squared of the radiation scalar, $|\Phi_2|^2$, is proportional to the radial component of
the Poynting vector, $S^R$.
Specifically, if we assume that $\Phi_2$ is calculated using the Kinnersley tetrad on a Kerr
background, from \cite{Teukolsky:1972my},
\beq
T^r_{{\rm EM} 0} = \frac{1}{2\pi} |\Phi_2|^2 = \frac{1}{\alpha} S^r,
\eeq
where the Boyer-Lindquist coordinate (areal) radius $r$ is adopted. As $r$ converges to the
numerical radial coordinate at large distances, and the lapse function
$\alpha \rightarrow 1$, we see that this is consistent with our definition of $\LEM$:
\beq
\LEM \equiv \lim_{R \rightarrow \infty} \oint R^2 S^R d\Omega
\eeq

\begin{figure}
  \includegraphics[trim=5mm 5mm 5mm 5mm,scale=.45]{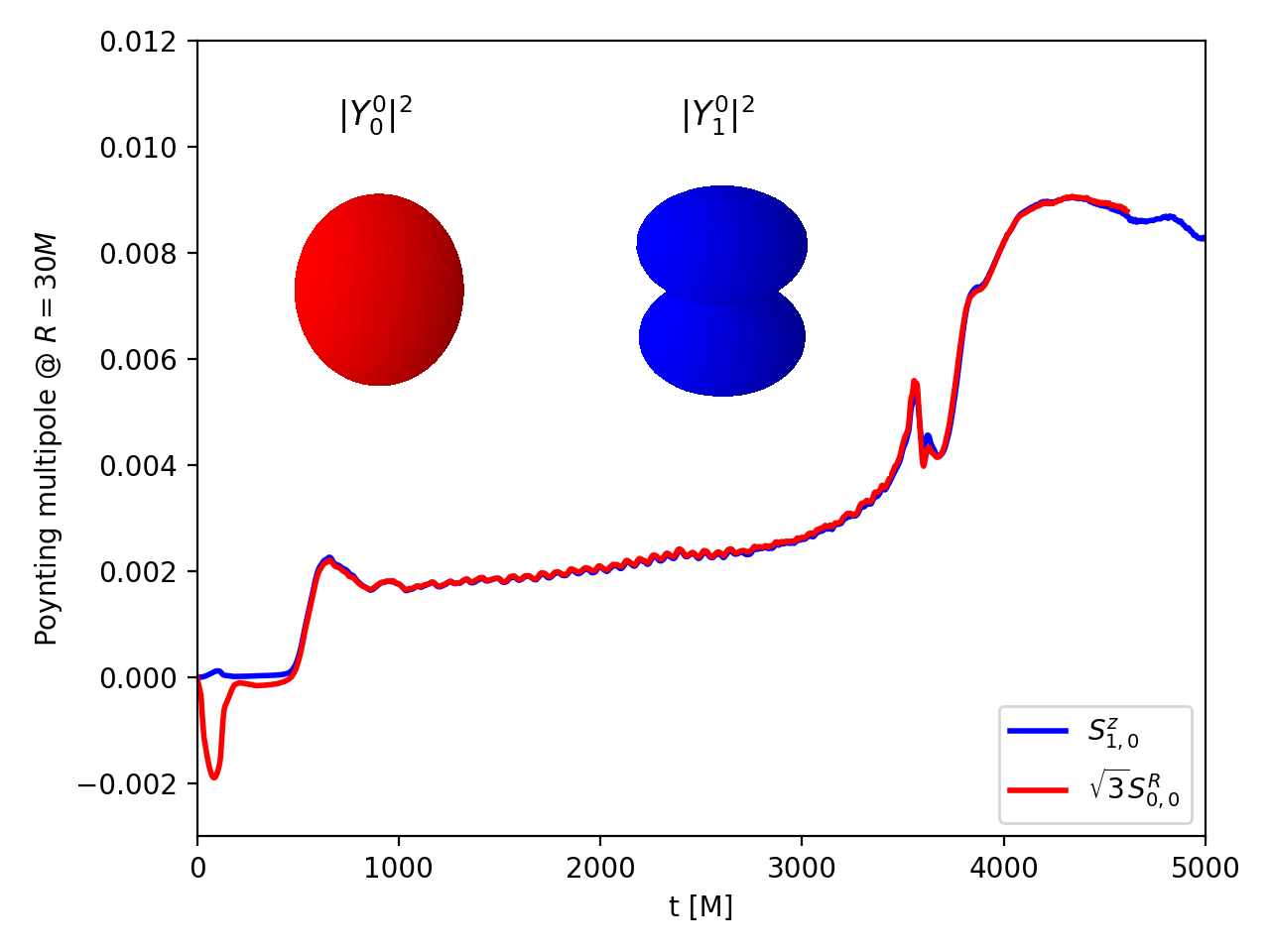}
  \caption{$\Szdom$ and $\sqrt{3} \Srdom$ extracted at $R = 30M$ for the $d = 14.4M$ configuration. The two
  quantities differ in the initial gauge relaxation pulse, but agree closely for the bulk of the signal
  beginning at $t \sim 450 M$, indicating that by this time the approximation $S^R \approx S^z \cos\theta$ holds.}
  \label{fig:Sz_10_V_Sr_00}
\end{figure}

In this case, we can relate the EM flux to the dominant $(\ell,m) = (1,0)$ spherical harmonic mode
of the Poynting vector used in this paper via

\begin{align}
\LEM &= \lim_{R \rightarrow \infty} \oint R^2 S^R d\Omega = \lim_{R \rightarrow \infty} 2 R^2 \sqrt{\pi} \Srdom \label{eq:LEM_as_Sr00}\\
     &\approx \lim_{R \rightarrow \infty} \oint R^2 S^z \cos\theta d\Omega = \lim_{R \rightarrow \infty} 2 R^2 \sqrt{\frac{\pi}{3}} \Szdom. \label{eq:LEM_as_Sz10_repeated}
\end{align}
This is the formula \eqref{eq:LEM_as_Sz10} used in our analysis.
In moving from \eqref{eq:LEM_as_Sr00} to \eqref{eq:LEM_as_Sz10_repeated}, we have assumed the Poynting flux is
dominated by emission along the polar ($z$) direction:
\beq
S^R \approx S^z \cos\theta \Rightarrow \Srdom \approx \frac{\Szdom}{\sqrt{3}}.
\eeq
This assumption is well-justified for the main part of the flux in the simulations presented here. For
example, in Fig.~\ref{fig:Sz_10_V_Sr_00}, we plot both $\Szdom$ and $\sqrt{3} \Srdom$ for the
$d = 14.4M$ configuration. The two signals differ in the initial gauge relaxation pulse (which is
therefore not $z$-dominated), but agree closely for the bulk of the signal beginning at $t \sim 500 M$.

\section{Resolution Tests and Convergence}
\label{apx:res_tests}

\begin{figure}
  \includegraphics[trim=5mm 5mm 5mm 5mm,scale=.60]{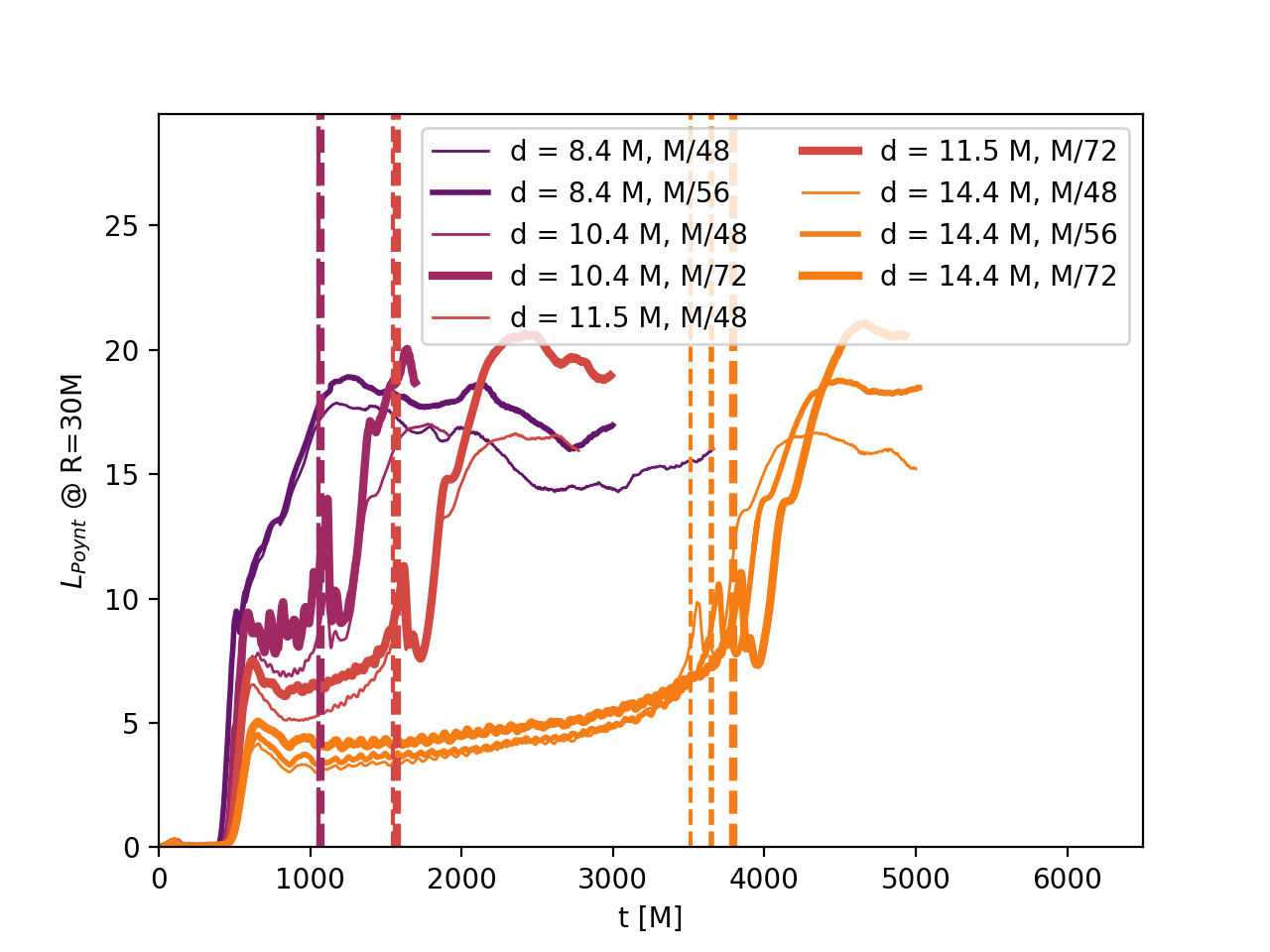}
  \caption{$\LEM$ for several configurations at basic ($dx = M/48$) and higher resolutions
  (denoted by thicker lines of the same color).}
  \label{fig:LEM_convergence}
\end{figure}

In Fig.~\ref{fig:LEM_convergence}, we look at the effect of resolution on the measured EM flux in
several of our configurations. 
It is evident that the general shape of the Poynting luminosity curve near feature (d) is
robust to changes in resolution, despite some sensitivity in the
quantitative level of the early rise and the post-merger plateau as
measured at this extraction radius.

\begin{figure}
  \includegraphics[trim=0mm 0mm 0mm 0mm,clip,scale=0.60]{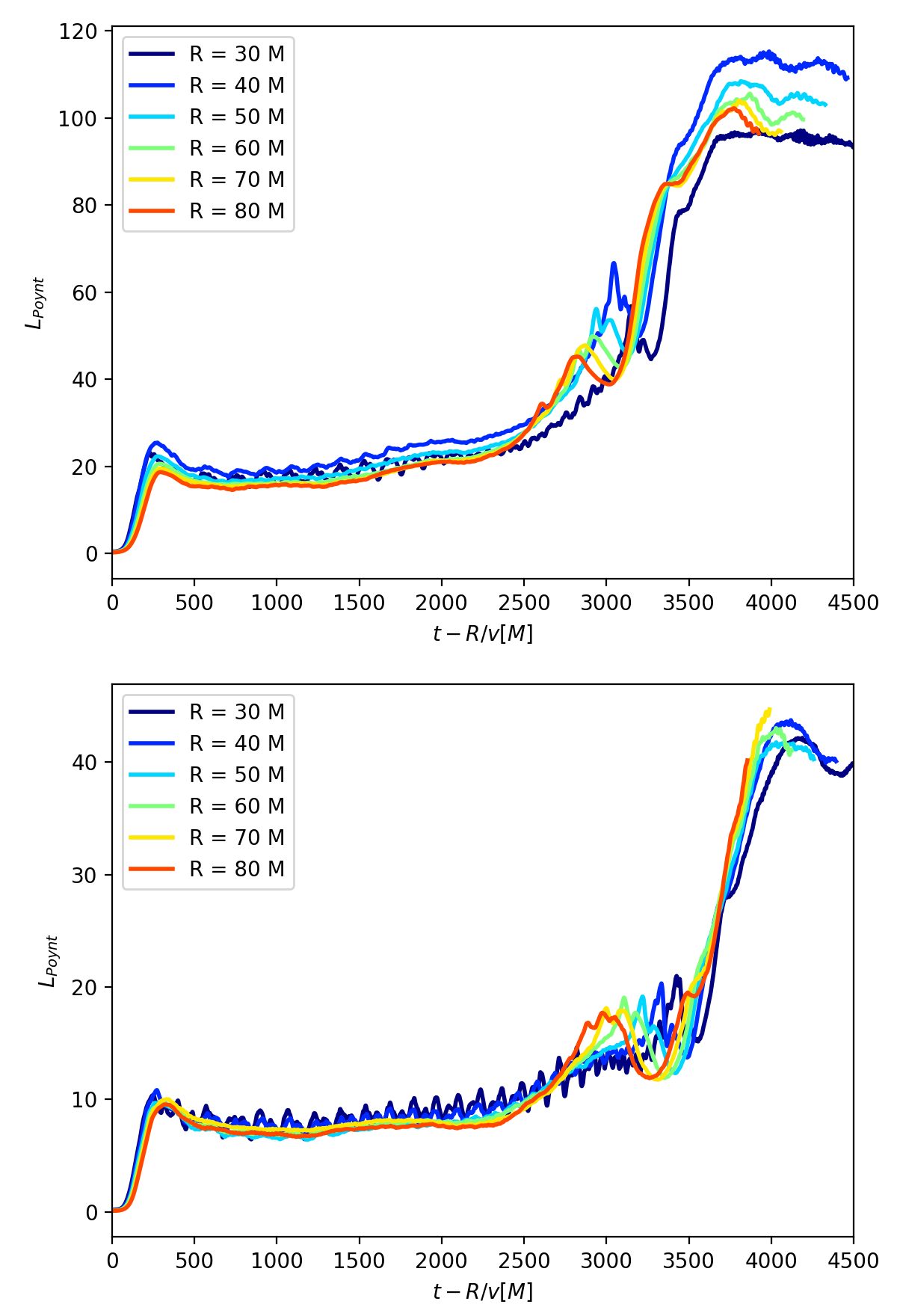}
  \caption{$\LEM$ for $d = 14.4M$ configuration, extracted at different radii, shifted in time consistent
  with a pulse speed $v = 0.07433 c$. The lower-resolution run (top) requires an additional multiplicative
  correction of $R^n$ where $n \approx 0.5$ for good alignment, while $n \approx 0.2$ is sufficient for the
  higher resolution.}
  \label{fig:LEM_allR_d14p4_multires}
\end{figure}

In Fig.~\ref{fig:LEM_allR_d14p4_multires}, we concentrate on one of the physical cases, $d = 14.4M$,
and show $\LEM$ calculated across several extraction spheres, $R/M \in
\{30, 40, 50, 60, 70,80\}$. We time-shift the different data sets
using the initial ambient Alfv\'{e}n speed $\valf = 0.07433$, which
serves to align the initial rise in Poynting flux to the inspiral
level. Note that for the least-resolved case ($h_{\rm f} = M/48$), the
measured luminosity drops with increased extraction radius $R$; while
some dissipation of Poynting flux is possible, the lower panel shows
that most of the effect vanishes for higher resolution ($h_{\rm f} =
M/72$), pointing to numerical dissipation as a major cause.
The shape of feature (d) does vary with the extraction radius, softening as the extraction
radius increases.

\section{Converting from Geometric to Gaussian/CGS Units}
\label{apx:geom_to_CGS}

The initial plasma configuration for the canonical field case was chosen so that $\rho/(b^2/2) = 200$
far from the strong-field regions.
Given that $\rho$ is a matter density, there has to be some conversion for this to make sense.

In Gaussian units, the fluid and magnetic energy densities are
\begin{equation*}
u_{\rm fluid} = \rho c^2,\; u_{\rm magnetic} = \frac{B^2}{8 \pi} = \frac{b^2}{2}.
\end{equation*}
Thus the ratio of the two is
\beq
\umagofluid \equiv \frac{u_{\rm magnetic}}{u_{\rm fluid}} = \frac{B^2}{8 \pi \rho c^2}.
\eeq
Then to get the field strength $B$ given a specified fluid density $\rho$ and energy ratio $\umagofluid$,
\begin{align}
B^2 &= 8 \pi \rho c^2 \umagofluid \nonumber \\
    &= 720 \pi \times 10^6 \umagofluid \rhothir \, \UNITg \, \UNITcm^{-1} \, \UNITs^{-2} \nonumber \\
\Rightarrow B &= \sqrt{\frac{36 \pi}{5} \rhothir \umagofluid} \times 10^4 \, \UNITG, \label{eq:convert_B}
\end{align}
where we define
$\rhothir \equiv \rho/(10^{-13} \, \UNITg \, \UNITcm^{-3})$.

Note that the expression \eqref{eq:LEM_rhodep} is in standard geometric code units, where
$G = c = 1$. To convert to dimensionful units, we must multiply by a factor $G^2/c$. Expressing
$\rho_0$ and $\MTOT$ in cgs units, this factor is approximately
$1.483 \times 10^{-25} \, \UNITg^{-2} \, \UNITcm^4 \, \UNITs^{-2}$. That is, we can rewrite
\eqref{eq:LEM_rhodep} as
\begin{align*}
\LEM(t) = 1.483 &\times 10^{-25} \left( \frac{\rho_0}{1 \, \UNITg \, \UNITcm^{-3}} \right) \left( \frac{\MTOT}{1 \, \UNITg}\right)^2\\
                & \times F(t;\epsilon_0,\umagofluid_0) \UNITerg \, \UNITs^{-1}.
\end{align*}

If instead, we scale with our canonical density
$\rho_0 = 10^{-13} \, \UNITg \, \UNITcm^{-3}$, and
a total system mass of $\MTOT = 10^8 \MSun = 1.989 \times 10^{41} \, \UNITg$, we find
\beq
\LEM(t) = 5.867 \times 10^{44} \rhothir M_8^2 F(t;\epsilon_0,\umagofluid_0) \, \UNITerg \, \UNITs^{-1},
\eeq
where we define
$\rhothir \equiv \rho_0/(10^{-13} \, \UNITg \, \UNITcm^{-3})$
and $M_8 \equiv \MTOT/10^8 \MSun$. As shorthand, we call this numerical factor $\eta_{\rm cgs}$:
\beq
\eta_{\rm cgs} \equiv 5.867 \times 10^{44} \rhothir M_8^2.
\label{eq:code_to_cgs}
\eeq

\end{document}